\documentclass[12pt]{article}

\usepackage{color} 

\usepackage{epsfig}
\usepackage{amssymb}
\usepackage[round]{natbib}

\newcommand{\bea}{\begin{eqnarray}}
\newcommand{\eea}{\end{eqnarray}}

\newcommand{\cip}{\perp\!\!\!\perp\,}

\newtheorem{defi}{Definition}

\newtheorem{corol}[defi]{Corollary}

\newtheorem{prop}[defi]{Proposition}

\begin{document}

\author{Stijn Vansteelandt \\[1ex] 
\textit{Department of Applied Mathematics, Computer Sciences and Statistics} \textit{\ }%
\\
\textit{Ghent University, Belgium} \and 
and Vanessa Didelez \\
\textit{School of Mathematics} \textit{\ }%
\\
\textit{University of Bristol, U.K.} 
}
\title{
Robustness and efficiency of covariate adjusted linear instrumental variable estimators }
\date{}

\maketitle

\bigskip \setlength{\parindent}{0.3in} \setlength{\baselineskip}{24pt}

Two-stage least squares (TSLS) estimators and variants thereof are widely used to infer the effect of an exposure on an outcome using instrumental variables (IVs). 
They belong to a wider class of two-stage IV estimators, which are based on fitting a conditional mean model for the exposure, and then using the fitted exposure values along with the covariates as predictors in a linear model for the outcome. We show that standard TSLS estimators enjoy greater robustness to model misspecification than more general two-stage estimators. However, by potentially using a wrong exposure model, e.g.\ when the exposure is binary, they tend to be inefficient. In view of this, we study double-robust G-estimators instead. These use working models for the exposure, IV and outcome  but only require correct specification of either the IV model or the outcome model to guarantee consistent estimation of the exposure effect. As the finite sample performance of the locally efficient G-estimator can be poor, we further develop G-estimation procedures with improved efficiency and robustness properties under misspecification of some or all working models. Simulation studies and a data analysis demonstrate drastic improvements, with remarkably good performance even when one or more working models are misspecified.


\section{Introduction}

An enormous body of research has developed in the econometrics and biostatistics literatures on how to assess the causal effect of an exposure $X$ on an outcome $Y$ in the presence of confounding by unobserved variables $U$, when a {\em vector of instrumental variables} $Z$ (IVs) is available (see e.g. Bowden and Turkington, 1985; Robins, 1994; Angrist et al., 1996; Greenland, 2000; Wooldridge, 2002;   Hern\'an and Robins, 2006; Didelez and Sheehan, 2007).
It is therefore not surprising that a variety of competing approaches
have been put forward. 
A simple and popular method is two-stage least squares (TSLS) estimation where, in the first stage, the exposure is predicted based on an ordinary least squares regression of the exposure on the IVs and covariates; in the second stage, the outcome is regressed on the predicted exposure and covariates via ordinary least squares regression, and the exposure coefficient is taken as the final IV estimator of the desired causal effect. 
The simplicity of this approach has encouraged the development of other two-stage estimators, which are obtained along the same lines, but employ possibly nonlinear regressions in the first or second stage (see e.g.\ Mullahy, 1997; or review in Didelez et al., 2010). Variations on two-stage estimators, that we do not consider in much detail here, include limited information maximum likelihood (LIML; Bowden and Turkington, 1985, chapter 4; or see Anderson 2004 for an historic account), Bayesian (Kleibergen and Zivot, 2003) or control function approaches (Wooldridge, 2002, chapter 6).
In view of their increasing popularity, we will contrast the two-stage approaches with G-estimators under structural mean models. These do not rely on separate fitting of a first stage model (Hern\'an and Robins, 2006) and have connections to Generalized Method of Moments (GMM) estimation (Clarke and Windmeier, 2010). 


It is natural to ask how the various IV methods compare with regard to their efficiency as well as robustness under various types of model misspecification. In particular, two-stage methods appear to rely on a correct exposure model but may in a variety of situations be consistent even if this is misspecified, or not be efficient even if the exposure model is correct. Moreover, in the presence of covariates, one can investigate the role of models  for the relation between covariates on the one hand, and exposure, outcome or instruments on the other hand.
Efficiency is relevant in this context as methods that are robust towards misspecification of some model assumptions will likely not be as efficient as methods exploiting a correctly specified  model. Moreover, one can ask whether including covariates (when there is the choice) typically leads to efficiency gains (as noticed in an IV setting by Fisher-Lapp and Goetghebeur, 1999) and whether there is a trade-off with robustness.

In this article we investigate these questions formally in the context of linear IV models; these are introduced in Section \ref{sec:model}. We then focus on two-stage IV estimators where we consider arbitrary (possibly nonlinear) conditional mean models for the exposure 
 combined with arbitrary linear conditional mean models for the outcome (Section \ref{sec:ts}). 
We study their efficiency, and their bias under misspecification of the exposure and/or outcome model. 
In Section \ref{sec:exp} we find a subclass of two-stage IV estimators to enjoy robustness against misspecification of the exposure model. We moreover derive the locally efficient IV-estimator which does not rely on correct specification of an exposure model, and find it to equal a specific two-stage estimator in some cases, but not in general.
As addressed in Section \ref{sec:outc}, a further subclass of IV estimators is double-robust: consistent if either a model for the main effect of covariates on the outcome or a model for the distribution of the IV, given covariates, is correctly specified, but not 
necessarily both (Okui et al., 2012). Interestingly, the TSLS estimator enjoys double robustness, but only w.r.t.\ a linear model for the conditional mean of the IV given covariates (Robins, 2000). Moreover, it is sometimes inefficient relative to other double-robust estimators, for instance when the exposure obeys a nonlinear model or when the exposure effect depends on covariates. 

In our simulation studies the locally efficient double-robust IV estimator (Robins, 1994) outperforms the TSLS estimator when based on correctly specified exposure and outcome models, but performs much worse otherwise. In view of this, we develop two adaptive estimation procedures in Section \ref{sec:impro}. 
The first makes use of empirical efficiency maximisation (Rubin and van der Laan, 2008) which is designed to maximise precision even  under misspecification of the exposure and outcome model, 
and results in drastic efficiency gains when the model for the distribution of the IV, given covariates, is correctly specified.
The second makes use of bias-reduced double-robust estimation (Vermeulen and Vansteelandt, 2015), which is designed to minimise bias even under additional misspecification of the IV distribution. Simulation studies confirm the bias-reduction and moreover demonstrate favourable performance regarding efficiency.

Proofs of the various propositions and corollaries in the article can be found in Appendix A of the Supplemental Materials.

\section{Linear instrumental variable models}\label{sec:model}

Let $Z$ be a vector of IVs for the effect of a scalar exposure $X$ on a scalar outcome $Y$, conditional on a vector of observed covariates $C$. 
In analogy to Didelez and Sheehan (2007) 
but extending the definition to account for covariates (see also Pearl, 2009, p.248), we formalise this by the assumptions that $Z$  is (a) associated with $X$ conditional on $C$, (b) independent of $Y$, conditional on $X,U$ and $C$, and (c) independent of $U$, conditional on $C$; here, $U$ is a (set of) latent variable(s) such that $(U,C)$ would be sufficient to control for confounding of the effect of $X$ on $Y$ were $U$ observable.
This formalisation of an IV is close to, but allows for greater flexibility than that in the econometric literature on IVs (Wooldridge, 2002), where assumptions are usually in terms of no correlation instead of independence. In the causal inference literature, conditions (b) and (c) are often alternatively formalised in the assumption (b')  that  $Y_x\cip Z|C$ (Robins, 1994),
with  $Y_x$ denoting the counterfactual outcome that would be observed when setting $X$ to $x$. 
The latter formulation avoids explicit reference to any specific unobserved confounders.

We start by briefly addressing the relationship of different formulations of linear IV models. 
Consider first the following model for the conditional mean of the outcome:
\begin{equation}\label{2sls}
E\left(Y|X,Z,U,C\right)=\omega(C,U)+m_{}(C;\psi^*)X.
\end{equation}
Here, $\omega(C,U)$ is an unknown (i.e., unspecified) function of measured and unmeasured covariates.
The term $m_{}(C;\psi)$ is a known function of observed covariates, smooth in $\psi$, and $\psi^*$ is an unknown finite-dimensional parameter, e.g. $m_{}(C;\psi)=\psi$ or $m_{}(C;\psi)=\psi^T C$, where 
with a slight abuse of notation, the vector $C$ includes 1 to allow for a main effect. When $m_{}(C;\psi)$ is parameterised such that 
$m_{}(C;\psi)=0$ when $\psi=0$, as in the previous examples, we have that $m_{}(C;\psi^*)$ captures the exposure effect of interest, i.e.\ we regard $\psi^*$ as the target `causal' parameter.  In particular,
\begin{eqnarray}
m_{}(C;\psi^*)X&=&E\left(Y|X,Z,U,C\right)-E\left(Y|X=0,Z,U,C\right)\nonumber\\
&=&E\left(Y|X,Z,U,C\right)-E\left(Y_0|X=0,Z,U,C\right)\nonumber\\
&=&E\left(Y|X,Z,U,C\right)-E\left(Y_0|X,Z,U,C\right)\nonumber\\
&=&E\left(Y|X,Z,C\right)-E\left(Y_0|X,Z,C\right), \label{ce}
\end{eqnarray}
which encodes the additive effect on the outcome of setting the exposure to zero in a subgroup of individuals with exposure $X$, IV $Z$ and covariates $C$. In the above derivation,
the second equality follows by the consistency assumption that $Y=Y_0$ in subjects with $X=0$, the third from the fact that $U$ and $C$ are sufficient to control for confounding of the effect of $X$ on $Y$, and the fourth by the fact that the left-hand side does not involve $U$.
Note that many IV methods assume a (linear) structural equation for the outcome $Y$ which is more restrictive but implies the above (\ref{2sls}) for $\omega(C,U)$ equal $\beta^{*T} C+U$.

The model defined by restriction (\ref{ce}), i.e.
\begin{eqnarray}
E(Y|X,Z,C)=E(Y_0|X,Z,C)+m_{}(C;\psi^*)X,
\label{lsmm}
\end{eqnarray}
is called a linear or additive  structural mean model (Robins, 1994). 
Together with the IV assumptions (a) and (b') it can be regarded as the substantive model of interest, as it merely parameterizes the exposure effect of interest. 
That the exposure effect does not involve $Z$ (or equivalently, that $Z$ does not appear on the right-hand side of (\ref{2sls}), but is included on the left hand side)
is known as `no effect modification' by $Z$ (Hernan and Robins, 2006; Clarke and Windmeier, 2010). It is this assumption which ultimately allows inference exploiting the IV $Z$ as we will see below. While it can be motivated by the additivity in (\ref{2sls}), it is often made in its own right avoiding explicit reference to $U$ and hence allowing greater generality. In other words,  models (\ref{2sls}) and (\ref{lsmm}) differ in their assumptions on unobservables, but they essentially impose the same restrictions on the observed data law under the IV assumptions (see the Supplementary Materials for details).  We therefore use the same notation, $\mathcal{M}$, throughout to denote both  IV models.

\section{Two-stage estimation}\label{sec:ts}

Model $\mathcal{M}$ cannot be fitted directly as $Y_0$ (resp.\ $U$) is unobserved. Two-stage approaches exploiting the IV $Z$ use the following restriction implied by $\mathcal{M}$:
\begin{equation}\label{odmodel}
E\left(Y|Z,C\right)=\omega(C)+m_{}(C;\psi^*)E(X|Z,C),\end{equation}
for $\omega(C)\equiv E\left(Y_0|C\right)$ (or $\omega(C)\equiv E\left\{\omega(C,U)|C\right\}$ if we start with (\ref{2sls})). 
When $C$ is high-dimensional (e.g.\ continuous or discrete with several components), the above cannot be fitted non-parametrically and additional modelling assumptions are needed to obtain estimators of $\psi^*$ with adequate performance in moderate sample sizes. Equation (\ref{odmodel}) suggests postulating two additional models, one for $E(X|Z,C)$ and one for $\omega(C)$, and thereby lays the basis of two-stage estimation procedures. 

In the first stage, a parametric model $\mathcal{A}_x$ is postulated for the exposure, i.e. 
\begin{equation}\label{modx}
E\left(X|Z,C\right)=m_x(Z,C;\alpha^*),
\end{equation}
where $m_x(Z,C;\alpha)$ is a known function of instruments and covariates, smooth in $\alpha$ and $\alpha^*$ is an unknown finite-dimensional parameter. An obvious choice would be a linear 
or logistic regression model (e.g., $m_x(Z,C;\alpha)=\mbox{expit}(\alpha_z^TZ+\alpha_c^TC)$). 
The second stage model supplements that structural model $\mathcal{M}$ with a parametric model $\mathcal{A}_y$ for the main effects of covariates on the outcome:
\begin{equation}\label{omega}
\omega(C)=m_y(C;\beta^*),
\end{equation}
where $m_y(C;\beta)$ is a known function of covariates, smooth in $\beta$ and $\beta^*$ is an unknown finite-dimensional parameter.


In the remaining sections, we will highlight results and computations that are specific to the following common choices for the models. We denote the model with a constant causal effect as
\begin{equation}
\mathcal{M}^{const}: \quad m(C;\psi)=\psi. \label{m.const}
\end{equation}
This is to be contrasted with the more general case allowing effect modification by some of the covariates in $C$.
Further, we will pay special attention to linear models for exposure and covariates
\begin{eqnarray}
\mathcal{A}^{lin}_x: &\quad& m_x(Z,C;\alpha)=\alpha_z^TZ+\alpha_c^T C \label{Ax.lin} \\
\mathcal{A}^{lin}_y: &\quad& m_y(C;\beta)=\beta^T C, \label{Ay.lin}
\end{eqnarray}
where it will often be important that the vector of covariates $C$ is indeed exactly the same in both (\ref{Ax.lin}) and (\ref{Ay.lin}). Note that model $\mathcal{A}^{lin}_x$ is quite general. In particular, it allows for covariate-instrument interactions by letting $Z$ equal $Z^*C$ for some IV $Z^*$.
More generally, the two models could use the covariates in a different way, e.g.\ with $V$ a scalar component of $C$, we may have $\alpha_1+\alpha_2\log V$ in modification of (\ref{Ax.lin}) combined with $\beta_1+\beta_2V+\beta_3V^2$ in modification of (\ref{Ay.lin}).

\subsection{Consistency of general two-stage estimators}\label{sec:2stage}

The model defined by all three types of restrictions, i.e.\ $\mathcal{M}\cap\mathcal{A}_x\cap\mathcal{A}_y$, implies a standard conditional mean model (Chamberlain, 1987) of the form
\begin{equation}\label{mody}
E\left(Y|Z,C\right)=m_y(C;\beta^*)+m_{}(C;\psi^*)m_x(Z,C;\alpha^*).
\end{equation}
For instance, the models $\mathcal{M}^{const}, \mathcal{A}^{lin}_x, \mathcal{A}^{lin}_y$ imply the linear regression model
\[
E\left(Y|Z,C\right)={\beta^*}^T C+\psi^*(\alpha_z^{*T}Z+{\alpha_c^*}^TC).
\]
Likewise, the logistic exposure model combined with a linear outcome model implies 
\[
E\left(Y|Z,C\right)={\beta^*}^T C+\psi^*\mbox{expit}(\alpha_z^{*T}Z+{\alpha_c^*}^T C).
\]
A general two-stage procedure is thus obtained by fixing $\alpha^*$ at some estimate $\hat{\alpha}$ obtained from fitting model (\ref{modx}) 
and then fitting model (\ref{mody})  with $\alpha^*$ substituted by $\hat \alpha$ using standard regression techniques at each stage.
Two-stage estimation thus arises very naturally in an IV context. However, as we will show, it is not generally consistent
under misspecification of model $\mathcal{A}_x$ or $\mathcal{A}_y$,
and not necessarily efficient even when both models are correct;
we would therefore not generally recommend this approach. We cover two-stage estimation  here because  some special cases, in particular the popular TSLS, exhibit greater robustness and efficiency, which we will later compare with other methods that are robust by design.

Recall that all Consistent Asymptotically Normal (CAN) estimators for $\alpha^*$ in $\mathcal{A}_x$ are asymptotically equivalent to the solution to an estimating equation of the form
\begin{eqnarray}
0&=&\sum_{i=1}^n e_x(Z_i,C_i)\left\{X_i-m_x(Z_i,C_i;\alpha)\right\}\label{2sx}
\end{eqnarray}
for index functions $e_x(Z,C)$ of the dimension of $\alpha$. For example, when $\mathcal{A}^{lin}_x$ is fitted using ordinary least squares estimation we have $e_x(Z,C)=(Z^T,C^T)^T$ in (\ref{2sx}).
Similarly, for a given $\alpha^*$, all CAN estimators for $\beta^{*},\psi^{*}$ in model (\ref{mody}) are asymptotically equivalent to the solution to an estimating equation of the form
\begin{eqnarray}
0&=&\sum_{i=1}^n e_y(Z_i,C_i)\left\{Y_i-m_y(C_i;\beta)-m_{}(C_i;\psi)m_x(Z_i,C_i;\alpha^*)\right\}\label{2sy}
\end{eqnarray}
for index functions $e_y(Z,C)$ of the dimension of $(\beta^T,\psi^T)^T$. 
Under $\mathcal{A}^{lin}_y$ and $\mathcal{M}^{const}$,   using ordinary least squares estimation amounts to choosing $e_y(Z,C)=({C}^T,\alpha^{*T}_z Z+\alpha_c^{*T} C)^T$ in (\ref{2sy}).
The estimators $\hat{\beta}$ for $\beta^*$ and $\hat{\psi}$ for $\psi^*$ resulting from substituting $\alpha^*$  by $\hat \alpha$ in (\ref{2sy}) are also sometimes called `plug-in' estimators; they are still CAN, as the following proposition asserts.

\begin{prop} \label{TS.can} {\em Two-stage estimator is CAN}\\
\setlength{\parindent}{0.3in} \setlength{\baselineskip}{24pt}
The two-stage IV estimator of the causal parameter $\psi^*$  obtained by fixing $\alpha^*$ at the consistent estimator $\hat{\alpha}$ obtained by solving (\ref{2sx}) for some conformable index function $e_x(Z,C)$, and next fitting model (\ref{mody}) by solving (\ref{2sy}) for some conformable index function $e_y(Z,C)$ (with $\alpha^*$ substituted by $\hat{\alpha}$), is consistent and asymptotically normal under $\mathcal{M}\cap\mathcal{A}_x\cap\mathcal{A}_y$.
\end{prop}
\vspace*{-.35cm}
Proof: this follows from the general theory of M-estimation under standard conditional mean models (Stefanski and Boos, 2002). $\square$
  
Although two-stage estimators are also sometimes used in nonlinear models for the outcome (e.g. logistic regression models and proportional hazard models), Proposition \ref{TS.can} relies on the identity (\ref{odmodel}) and is difficult to justify outside the realm of such additive causal models as (\ref{2sls}) or (\ref{lsmm}) (see also Didelez, Meng and Sheehan, 2010; Vansteelandt et al., 2011); an exception occurs in the context of additive hazard models as discussed in Tchetgen Tchetgen et al.\ (2015).

\subsection{Two-stage least-squares estimation (TSLS)}

Among IV methods, TSLS takes a prominent place and provides an apparent two-stage estimator (Wooldridge, 2002). The principle of TSLS is that all `endogenous' exposures (those that are confounded, i.e.\ dependent on $U$) are replaced by their linear projections on all `exogenous' variables (these are the IVs, covariates, and possible other unconfounded  exposures in the outcome model). 
As the name suggests,  TSLS is equivalent to explicit two-stage estimation because the linear projections are equivalent to fitting a linear first stage model with ordinary least squares, and these can then be plugged into the second stage model, again fitted by least-squares. For the equivalence it is, however, important to use the {\em implied} first stage model, i.e.\ a linear model for $X$ (and other endogenous variables) given {\em all} exogenous variables as determined by the choice of IVs and $\mathcal{A}_y$. In the following we refer to this procedure as  {\em Standard TSLS}, to distinguish it from \textit{Plug-In TSLS}, which also uses least squares in the first and second stage, but under possibly more general linear models $\mathcal{A}_x,\mathcal{A}_y$, e.g.\ with different transformations of the covariates.


We illustrate Standard TSLS with an example. Consider the case where $C=(1,V)^T$ with $V$ a scalar, 
$m_{}(C;\psi)X=\psi_1X+\psi_2XV$ and $m_y(C;\beta)=\beta_0+\beta_1V+\beta_2V^2$. 
There are two `endogenous' variables, $X$ and $XV$, as these both depend on $U$. For identification it is necessary that there are at least as many instruments as endogenous variables; hence, two instruments are needed, which could be $Z$ and $ZV$. 
The linear projections would be of $X$ and $XV$ each on all of $Z$, $ZV$, $V$ and $V^2$. It follows that the implied first stage models are  $E(X|Z,C)=\alpha_0+\alpha_1Z+\alpha_2ZV+\alpha_3V+\alpha_4V^2$ and  $E(XV|Z,C)=\alpha'_0+\alpha'_1Z+\alpha'_2ZV+\alpha'_3V+\alpha'_4V^2$ where it is assumed that the coefficients of the instruments in the projections are non-zero (more precisely, that  the matrix with first row $\alpha_1,\alpha_2$ and second row $\alpha'_1,\alpha'_2$ has full rank); the latter is a more specific version of assumption (a).

\subsection{Efficiency of two-stage IV estimation}\label{sec:effTS}

Even when model $\mathcal{M}\cap\mathcal{A}_x\cap\mathcal{A}_y$ is correctly specified, two-stage IV estimators are not generally efficient under this model  because they are based on {\it separately} fitting the exposure model and the outcome model. Since the parameter $\alpha^*$ indexing the exposure model also appears in the outcome model (\ref{mody}), simultaneous fitting of the exposure and outcome model may sometimes result in more efficient estimators of $\psi^*$ under  model $\mathcal{M} \cap \mathcal{A}_x \cap \mathcal{A}_y$ (this is outlined in Appendix A of the Supplementary Materials as part of the proof of Proposition \ref{eff.tsls}).
However, Proposition \ref{eff.tsls} shows that efficiency {\em is} achieved for the Standard TSLS estimators, despite their apparent two-stage nature, under 
model $\mathcal{M}^{const}\cap\mathcal{A}_x^{lin}\cap\mathcal{A}_y^{lin}$, provided that the conditional variance-covariance matrix $\mbox{Cov}((X,Y)^T|Z,C)$ is constant in $Z,C$.

\begin{prop} \label{eff.tsls} {\em Efficiency of Standard TSLS estimators}\\
\setlength{\parindent}{0.3in} \setlength{\baselineskip}{24pt}
When the conditional variance-covariance matrix $\mbox{Cov}((X,Y)^T|Z,C)$ is constant in $Z,C$, then the Standard TSLS estimator of $\psi^*$ is semi-parametric (locally) efficient in 
model $\mathcal{M}^{\rm const}\cap\mathcal{A}_x^{lin}\cap\mathcal{A}_y^{lin}$.
\end{prop} 


Proposition \ref{eff.tsls} does not immediately extend to more general two-stage IV estimators. In particular, two-stage estimators (including TSLS estimators) may be inefficient when the exposure effect depends on covariates (i.e. when $m_{}(C;\psi^*)=\psi^{*T}C$), even when the exposure and outcome model are fitted using ordinary least squares regressions. 
For TSLS estimators, this can be intuitively seen because for instance when $m_{}(C;\psi^*)=\psi^{*}_0+\psi^{*}_1V$ for $C=(1,V)^T$, TSLS is based on separate  least squares regressions of $X$ and $XV$ on $Z$ and $V$, without taking into account that the model for $X$ implies the model for $XV$, and without considering that the postulated models may be incompatible (e.g., even when the model for $X$ includes a main effect of $V$, the model for $XV$ may not allow for a main effect of $V^2$).
Two-stage estimators (including TSLS estimators) are moreover generally inefficient when the true exposure relation is nonlinear in $Z$ or $C$ (e.g. because it includes an interaction between $Z$ and components of $C$, or because it is of the logistic form), or when the outcome is dichotomous so that $\mbox{Cov}((X,Y)^T|Z,C)$ is not constant in $Z,C$. In particular, it may happen under certain data laws that the Standard TSLS estimator does not exist (more precisely, is not $\sqrt{n}$-consistent), even though other two-stage estimators with small variance exist. This is for instance the case when $E(X|Z,C)=Z-ZV$ for a scalar variate $V\in C$ which takes the values 0 and 1 with probability 1/2, independently of $Z$, and when furthermore $m_{}(C;\psi)X=\psi X$ and $m_y(C;\beta)=\beta_0+\beta_1V$. In that case, the implied first stage model would ignore the interaction between $Z$ and $V$ and thus result in $E(X|Z,C)=0$, thereby violating the necessary rank condition for TSLS estimators. In those cases, a Plug-In TSLS estimator based on a first stage model that includes main effects of $Z$, $V$ and their interaction, is indicated. 

\section{Estimation without reliance on an exposure model}\label{sec:exp}

It follows from the proof of Proposition \ref{eff.tsls} that simultaneous fitting of the exposure and outcome model may be needed in order to obtain a semi-parametric efficient estimator of $\psi^*$ in model $\mathcal{M}\cap\mathcal{A}_x\cap\mathcal{A}_y$. However, as this estimator may lack robustness against misspecification of $\mathcal{A}_x$, we consider and recommend more robust procedures in this and subsequent sections.
Before we address an estimation procedure that does not require an exposure model  we note that certain Plug-in TSLS estimators of $\psi^*$ in model $\mathcal{M}^{const}$
enjoy robustness against misspecification of $\mathcal{A}_x$, in particular the Standard TSLS estimator (Robins, 2000; Wooldridge, 2002), even though they were not designed to this end. 
The basic, and maybe somewhat paradoxical, rationale is that for particular choices of $\mathcal{A}_x$ it turns out that the estimating equations remain valid even if $\mathcal{A}_x$ is misspecified while for other choices of $\mathcal{A}_x$ the estimating equations then lose their validity. 

We can see the robustness by noting that solving (\ref{2sy}) will be equivalent to solving 
\begin{equation}
\label{eelin2}
0=\sum_{i=1}^n e_y(Z_i,C_i)\left\{Y_i-m_y(C_i;\beta)-m_{}(C_i;\psi)X_i\right\}
\end{equation}
whenever the index function $e_x(Z,C)$ in (\ref{2sx}) for fitting the exposure model $\mathcal{A}_x$ includes the component 
\begin{equation}
\label{Ax_robust}
e_y(Z,C)m_{}(C;\psi),
\end{equation}
for all $\psi$ (or some full-rank linear transformation of it). This is because the fitting procedure for the exposure model then ensures that 
\begin{equation}
\label{x.robust}
\sum_{i=1}^n e_y(Z_i,C_i)m_{}(C_i;\psi)X_i=\sum_{i=1}^n e_y(Z_i,C_i)m_{}(C_i;\psi)m_x(Z_i,C_i;\hat{\alpha})
\end{equation}
for all $\psi$.
Estimating equation (\ref{eelin2}) no longer involves the exposure model. In particular, its unbiasedness is not dependent upon (correct) specification of an exposure model. For arbitrary functions $e_y(Z,C)$ of the dimension of $(\beta,\psi)$, the solution to equation (\ref{eelin2}) is thus a CAN estimator  of $(\beta^*,\psi^*)$ in model $\mathcal{M}\cap\mathcal{A}_y$, i.e., regardless of (correct) specification of model $\mathcal{A}_x$. 
In Section \ref{subsec:rob2s}, we use the above result to show that certain two-stage estimators exhibit robustness against misspecification of the exposure model. In Section \ref{subsec:robeff}, we derive the semi-parametric efficient estimator under model $\mathcal{M}\cap\mathcal{A}_y$ (i.e., the optimal index function $e_y(Z,C)$ in (\ref{eelin2})). 

\subsection{Robustness of two-stage estimators against misspecification of the exposure model}\label{subsec:rob2s}

Condition (\ref{x.robust}) is met by certain two-stage estimators in linear models that are fitted using ordinary least squares, as detailed below.

\begin{prop} {\em Robustness of Standard TSLS estimators against exposure model misspecification (see also Robins (2000) and Wooldridge (2002, Theorem 5.1))}\\
\label{prop:rob.2sls1}
\setlength{\parindent}{0.3in} \setlength{\baselineskip}{24pt}
The Standard TSLS estimator of $\psi^*$ is CAN under model $\mathcal{M}\cap\mathcal{A}^{lin}_y$.
This estimator is therefore still consistent when the implied exposure model $\mathcal{A}^{lin}_x$ is not a correct model for $E(X|Z,C)$. \end{prop}

While Proposition \ref{prop:rob.2sls1} is stated for Standard TSLS, it can be somewhat extended to other two-stage estimators as discussed in Appendix A of the Supplementary Materials. For instance,
robustness can be attained by ensuring that the exposure model is linear  and {\em minimally} includes the covariates in the outcome model. 
The above result does however, not extend to general two-stage estimators. It fails for plug-in estimators when the exposure effect depends on covariates, e.g.\ when $m_{}(C;\psi)=\psi^TC$,  or when the exposure model is nonlinear. For instance, consider fitting model $E(Y|Z,C)=\beta^{*T}C+m_x(Z,C;\alpha^*)\psi^{*T}C$ by ordinary least squares. Because ordinary least squares uses index function $e_y(Z,C)=(C^T,m_x(Z,C;\alpha^*)C^T)^T$, we have that $e_y(Z,C)m_{}(C;\psi)$ is then no longer contained in $e_x(Z,C)$ to ensure identity (\ref{x.robust}). 


\subsection{Local efficiency without exposure model}\label{subsec:robeff}


It is not necessary to rely on two stage estimators being `accidentally' robust towards misspecification of $\mathcal{A}_x$ (i.e.\ satisfying the conditions of Proposition \ref{prop:rob.2sls1}); we can instead estimate $\psi^*$ straightaway by solving an estimating equation of the form (\ref{eelin2}) yielding a CAN estimator under $\mathcal{M}\cap\mathcal{A}_y$ as stated below. Without relying on $\mathcal{A}_x$, we may however lose efficiency compared to two-stage estimation when $\mathcal{A}_x$ {\em is} correctly specified. Hence, in order to achieve greatest efficiency possible under $\mathcal{M}\cap\mathcal{A}_y$, the following Proposition \ref{prop:cany} also addresses the optimal choice of the index function $e_y(Z,C)$ in (\ref{eelin2}). 


\begin{prop} {\em Semi-parametric efficient CAN estimation under model $\mathcal{M}\cap\mathcal{A}_y$ \label{prop:cany}}\\
\setlength{\parindent}{0.3in} \setlength{\baselineskip}{24pt}
The IV estimator of the causal parameter $\psi^*$  obtained by solving (\ref{eelin2}) for some conformable index function $e_y(Z,C)$ is CAN under $\mathcal{M}\cap\mathcal{A}_y$; it does not rely on an exposure model. Moreover, all CAN estimators of $\psi^*$ in model $\mathcal{M}\cap\mathcal{A}_y$ are asymptotically equivalent to the solution of (\ref{eelin2}) for some conformable index function $e_y(Z,C)$.
A semiparametric locally efficient estimator of $(\beta^*,\psi^*)$ is obtained by choosing $e_y(Z,C)$ equal to 
\[e_{\rm y,opt}(Z,C)=\left(\begin{array}{cc} m_x(Z,C;\alpha^*)\partial m_{}(C;\psi^*)/\partial\psi\\
\partial m_y(C;\beta^*)/\partial\beta \end{array}\right)\mbox{Var}\left\{Y-m_{}(C;\psi^*)X|Z,C\right\}^{-1};\]
the efficiency is local in the sense that it depends on specification of a working model  $\mathcal{A}_x$, and is only attained when model $\mathcal{M}\cap\mathcal{A}_y\cap\mathcal{A}_x$ is correctly specified and the conditional variance $\mbox{Var}\left\{Y-m_{}(C;\psi^*)X|Z,C\right\}$ is consistently estimated (at faster than $n^{1/4}$ rate).
\end{prop}


To illustrate the above, let $m_x(Z,C;\alpha)=\mbox{expit}(\alpha_z^TZ+\alpha^{T}_cC)$, $m_y(C;\beta)=\beta^TC$ and $m_{}(C;\psi)=\psi^TC$. Then under homoscedasticity (i.e. $\mbox{Var}\left\{Y-m_{}(C;\psi^*)X|Z,C\right\}$ is constant)
a semi-parametric efficient estimator of $\psi^*$ in model $\mathcal{M}\cap\mathcal{A}_y$ is obtained by first estimating $\alpha^*$ using standard logistic regression, and next solving
\[0=\sum_{i=1}^n \left(\begin{array}{c}
\mbox{expit}(\hat{\alpha}_zZ_i+\hat{\alpha}^{T}_cC_i)C_i\\
C_i
\end{array}\right)(Y_i-\beta^TC_i-\psi^TC_iX_i);\] 
by the linearity of this equation in $\psi$, a semi-parametric (locally) efficient estimator is thus obtainable in closed form.

Wooldridge (2002) also studies semi-parametric efficiency under model $\mathcal{M}\cap\mathcal{A}_y$, but his results relate to the subclass of estimators obtained by solving (\ref{eelin2}) for $e_y(Z,C)$ linear in $Z$ and $C$. He shows that Standard TSLS estimators are efficient within this class under model $\mathcal{M}\cap\mathcal{A}_y$. Likewise, efficiency results for GMM estimators (Wooldridge, 2002) relate to a subclass of all estimators under model $\mathcal{M}\cap\mathcal{A}_y$ (namely, the solutions to arbitrary linear combinations, with constant coefficients, of a given number of unbiased estimating equations). It follows from the above expression for the efficient score that potentially more efficient estimators than TSLS estimators can be obtained under model $\mathcal{M}^{const}\cap\mathcal{A}_y$  
when the exposure is nonlinear in $Z$ or $C$, or when there is heteroscedasticity.

In Appendix A of the Supplementary Materials we compare more generally the efficiency of the approach in Proposition \ref{prop:cany} to that achieved using semi-parametric efficient joint estimation under a correct model for $\mathcal{A}_x.$
We find that the loss in efficiency by not relying on correct specification of an exposure model is low when the confounding is weak or under $\mathcal{A}^{lin}_x$ as implied by TSLS. Conversely, efficiency can potentially be gained when confounding is strong and e.g.\ $\mathcal{A}_x$ is not linear.

\section{Double-robust estimation}\label{sec:outc}


As the estimators of Proposition 4 do not rely on correct specification of an exposure model their validity is merely predicated upon correct specification of the structural model $\mathcal{M}$ in the absence of covariates (i.e., when $C$ is empty). When covariate adjustment is necessary, either because the IV assumptions are only satisfied conditional on covariates, or because of interest in effect heterogeneity, then they do rely on an outcome model, $\mathcal{A}_y$.  Misspecification of that model may then sometimes result in biased effect estimates. 

Robustness against misspecification of the outcome model $\mathcal{A}_y$ is achieved by further restriction to a subclass of estimators for $\psi^*$ obtained by solving (\ref{eelin2}) 
with $e_y(Z,C)$ an arbitrary vector function whose first $p$ components, with $p$ the dimension of $\psi$, equal
\[
e(Z,C)-E\left\{e(Z,C)|C\right\}
\]
for some arbitrary function $e(Z,C)$. Here, the conditional expectation $E\left\{e(Z,C)|C\right\}$ is calculated under a parametric model $\mathcal{A}_z$ defined by
\[
f(Z|C)=f(Z|C;\gamma^*),
\]
where $f(Z|C;\gamma)$ is a known density function, smooth in $\gamma$, and $\gamma^*$ is an unknown finite-dimensional parameter, which can be substituted by its maximum likelihood estimator $\hat{\gamma}$. 
For instance, when $Z$ is binary, we may assume that $P(Z=1|C)=\mbox{expit}(\gamma^{*T} C)$ and use standard logistic regression to estimate $\gamma^*$.
Further, let $\hat{\beta}$ be a consistent estimator of $\beta^*$ as obtained in the previous section. Then an estimator of $\psi^*$ can be obtained by solving 
\begin{equation}\label{eeg}
0=\sum_{i=1}^n \left[e(Z_i,C_i)-E\left\{e(Z_i,C_i)|C_i;\hat{\gamma} \right\}\right]\left\{Y_i-m_y(C_i;\hat{\beta})-m_{}(C_i;\psi)X_i\right\},
\end{equation}
for some conformable vector function $e(Z,C)$.
\begin{prop} \label{can.g-est}
{\em CAN estimation under model $\mathcal{M}\cap(\mathcal{A}_y\cup\mathcal{A}_z)$}\\
\setlength{\parindent}{0.3in} \setlength{\baselineskip}{24pt}
The IV estimator of the causal parameter $\psi^*$  obtained by solving (\ref{eeg}) for some conformable index function $e(Z_i,C_i)$ is CAN under $\mathcal{M}\cap(\mathcal{A}_y\cup\mathcal{A}_z)$. Moreover, all CAN estimators of $\psi^*$ in model $\mathcal{M}\cap(\mathcal{A}_y\cup\mathcal{A}_z)$ are asymptotically equivalent to the solution of (\ref{eeg}) for some conformable index function $e(Z_i,C_i)$.
\end{prop}
\vspace*{-.35cm}
Proof: see Robins (2000) and Okui et al. (2012).  $\square$

Because the solution to (\ref{eeg}) is a CAN estimator of $\psi^*$ when either working model $\mathcal{A}_z$ or $\mathcal{A}_y$ holds, in addition to the linear IV model $\mathcal{M}$, it has been called double-robust (Robins and Rotnitzky, 2001; Okui et al., 2012). The resulting estimators, which are also known as G-estimators (Robins, 1994), are especially attractive in studies where the law of $Z$ given $C$ is known as this guarantees robustness against misspecification of $\mathcal{A}_y$. Such knowledge, leading to correct specification of $\mathcal{A}_z$, is for instance given in randomized experiments where $Z$ denotes randomization, or in Mendelian randomization studies where the genetic instrument is often known to be independent of covariates $C$, in which case $E\left\{e(Z,c)|C=c\right\}$ can be consistently estimated as $n^{-1}\sum_{i=1}^n e(Z_i,c)$. 
Note that typical two-stage estimators fail to exploit such knowledge of the law of $Z$ given $C$.

\subsection{Robustness of two-stage estimators against misspecification of the exposure and outcome model}

The double-robustness property can be used to show that misspecification of the outcome model does not result in biased exposure effect estimates for the Standard TSLS estimator of $\psi^*$ when the IV  $Z$ happens to be linear in the covariates of the outcome model (or, in particular, independent of $C$) in the sense that $E(Z|C)=\gamma^{*T} C$ (Robins, 2000; Okui et al., 2012); this is interesting as Standard TSLS appears not to make use of $\mathcal{A}_z$.

\begin{prop} {\em Robustness of TSLS estimators against outcome model misspecification}\\
\label{prop:rob.2sls2}
\setlength{\parindent}{0.3in} \setlength{\baselineskip}{24pt}
The Standard TSLS estimator of $\psi^*$ in
model $\mathcal{M}$ with $m_{}(C;\psi)$ linear in $C$ is CAN under model $\mathcal{M}$ when $E(Z|C)=\gamma^{*T} C$.
\end{prop}

It follows from Proposition \ref{prop:rob.2sls2} that when $Z$ and $C$ are not independent, the robustness of the Standard TSLS estimator does not extend to general IVs, e.g. dichotomous IVs that obey a logistic regression model with main covariate effect $C$, nor to general two-stage estimators that involve nonlinear exposure models or effect heterogeneity (i.e. $m_{}(C;\psi^*)$ depending on $C$). 
It further follows from the proof of Proposition \ref{prop:rob.2sls2}  in the Supplemental Materials that the Plug-in TSLS estimator of $\psi^*$  is CAN under model $\mathcal{M}^{const}$ when the conditional mean $E\left\{m_x(Z,C;\alpha^*)|C\right\}$ is linear in $C$, and in the more general model $\mathcal{M}$ with $m_{}(C;\psi)$ linear in $C$ when $Z$ is independent of $C$, 
but not necessarily otherwise. Thus, when $E(Z|C)$ is linear in $V$ and $V^2$ (with $C=(1,V)^T$), then the Plug-in TSLS estimator will only be robust against outcome model misspecification when the outcome model includes the term $V^2$ (regardless of whether it is associated with the outcome).

\subsection{Efficiency of double-robust and TSLS estimators}

When choosing $e(Z,C)$ in (\ref{eeg})  one may want to consider the efficiency of the resulting estimator and ask whether it is worthwhile including covariates at all when there is the choice. To address this, we first recall how a a semi-parametric (locally) efficient estimator of $\psi^*$ under model $\mathcal{M}\cap\mathcal{A}_z$ is obtained. 
It follows from Robins (1994) (see also Okui et al., 2012)
that  $e(Z,C)$  in (\ref{eeg}) should then be equal to
\begin{equation}
\label{eeff}
e_{\rm opt}(Z,C)= \sigma^{-2}(Z,C)\frac{\partial m_{}(C;\psi^*)}{\partial\psi}\left[E(X|Z,C)-\frac{E\left\{\sigma^{-2}(Z,C)E(X|Z,C)|C\right\}}{E\left\{\sigma^{-2}(Z,C)|C\right\}}\right]\end{equation}
with $\sigma^2(Z,C)\equiv \mbox{Var}\left\{Y-m_{}(C;\psi^*)X|Z,C\right\}$. Since model $\mathcal{M}\cap\left(\mathcal{A}_z\cup\mathcal{A}_y\right)$ is less restrictive, this is also delivering the (locally) efficient estimator of $\psi^*$ in model $\mathcal{M}\cap\left(\mathcal{A}_z\cup\mathcal{A}_y\right)$. 
For instance, assuming that $E(X|Z,C)=\alpha^{*T}_{1}C+\alpha^{*T}_{2}ZC$ for scalar $Z$ and $C$, $m_y(C;\beta)=\beta^T C$ and $\sigma^2(Z,C)=\sigma^2$
for unknown parameters $\alpha^*_1,\alpha^*_2$ and $\beta^*$, we have
\begin{equation}\label{adaptive}
e_{\rm opt}(Z,C)= \sigma^{-2}\alpha^{*T}_2C\left\{Z-E(Z|C)\right\}.\end{equation} 
A locally efficient G-estimator may now be obtained by substituting $\alpha^*_2$ by the ordinary least squares estimator in the above expression, setting $\sigma^2$  to 1 (as it is just a proportionality constant), and next solving   (\ref{eeg}) for the resulting choice of $e(Z,C)=e_{\rm opt}(Z,C)$. These expressions suggest a way to optimally include covariates and, in a similar vein, to optimally combine multiple instruments (see e.g.\ Bowden and Vansteelandt, 2011).

In the special case where $Z\cip C$  and under $\mathcal{M}^{const}$, one has the choice of whether to adjust for $C$ at all. Consistent estimation can then also be achieved ignoring $C$. However, 
provided  $\mathcal{M}^{const}$ and working models for $E(Y-\psi^*X|C)$, $E(X|Z,C)$ and $\sigma^2(Z,C)$ are correctly specified, the covariate-adjusted analysis will then be at least as efficient in large samples as the unadjusted analysis.
While an efficiency gain is not generally guaranteed when these working models are misspecified, the following corollary demonstrates that it {\em can} (almost) always be guaranteed for the special case of Standard TSLS.

\begin{corol} \label{corol.tsls}
{\em Efficiency of covariate adjusted Standard TSLS estimators \\ }
\setlength{\parindent}{0.3in} \setlength{\baselineskip}{24pt}
When $Z\cip C$ and under $\mathcal{M}^{const}$, if $Y-\psi^*X$ is conditionally independent of $Z$ given $C$, then covariate adjustment does not increase (and usually reduces) the asymptotic variance  of the Standard TSLS estimator of $\psi^*$.
\end{corol}
\vspace*{-.35cm}
Proof: this follows as a special case of Proposition \ref{prop:rob.2sls3}; the proof is given in Appendix B of the Supplemental Materials. $\square$

The condition that $Y-\psi^*X$ be conditionally independent of $Z$ given $C$ in the above corollary can be regarded as a stronger version of the assumption of `no effect modification' by $Z$ common to IV models (Hernan and Robins, 2006; Clarke and Windmeier, 2010).

Fisher-Lapp and Goetghebeur (1999) also noticed that covariate adjustment is typically beneficial in a linear IV context; however, their results are specific to the case of partial compliance with full compliance in the control arm, where by design there is a corresponding interaction in the exposure model and where the IV model is specific to the treatment arm.

\subsection{Simulation study}\label{subsec:sim1}

\begin{figure}[htbp]
\centering
\includegraphics[width=\textwidth]{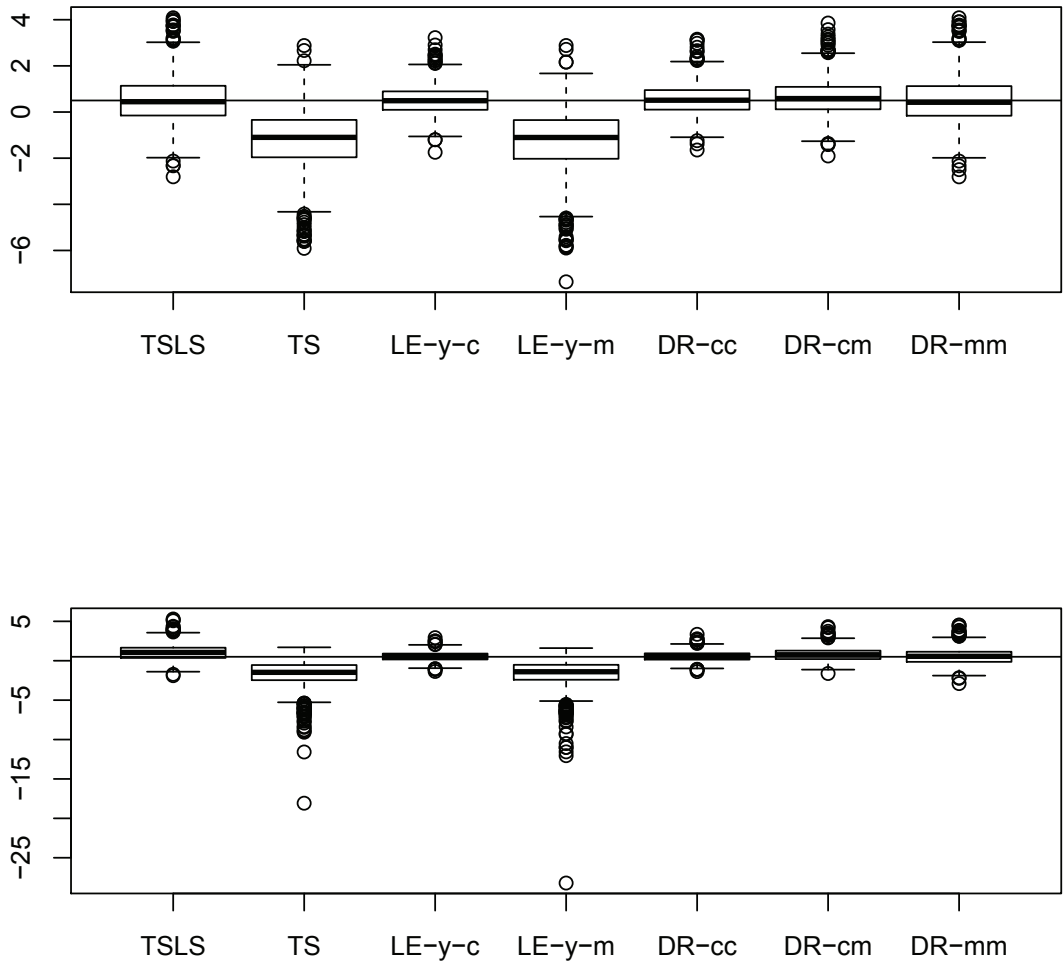}
\caption{Simulation results for binary exposure. Top: instrument independent of covariates; Bottom: instrument nonlinear in covariates.}
\label{fig:sim1:bp}
\end{figure}

To better appreciate the above results, we show empirical results from a small simulation study with $n=500$ in Figure \ref{fig:sim1:bp} (a more extensive simulation study follows in Section \ref{sec:sim}).
We generated mutually independent and standard normal covariates $U$ and $V$, $Z$ dichotomous with $P(Z=1|U,V)=0.27$, $X$ dichotomous with $P(X=1|Z,U,V)=\Phi(Z+U+V)$ and $Y$ normal with mean $0.5X-U-2V+V^2$. Assuming a linear IV model with $m_{}(C;\psi)=\psi$, a logistic model for the IV with $P(Z=1|C)=\mbox{expit}(\alpha_0+\alpha_1V)$  and $C=(1,V)^T$, we then evaluated the following estimators: 
\begin{description}
\item
(TSLS): the Standard TSLS using a misspecified outcome models in that it excludes $V^2$ and, implicitly, a misspecified linear exposure model; 
\item
(TS): the plug-in two-stage estimator with $P(X=1|Z,C)=\Phi(\alpha_zZ+\alpha_c^TC)$ (under which we obtained an average Nagelkerke pseudo-$R^2$ value of 0.24 and an F-value of 78) and $E(Y| Z,C)=\beta^TC+\psi\Phi(\hat{\alpha}_zZ+\hat{\alpha}_c^TC)$ which also wrongly excludes $V^2$; 
\item
(LE-y-c) and (LE-y-m): two locally efficient estimators under model $\mathcal{M}\cap\mathcal{A}_y$  as introduced in  Section \ref{sec:exp}; the first with correct outcome model, the second with incorrect outcome model (i.e. $m_y(C;\beta)=\beta^TC$), but both using the correct exposure model;
\item
(DR-cc, DR-cm and DR-mm): three double-robust semiparametric locally efficient estimators under model $\mathcal{M}\cap(\mathcal{A}_y\cup\mathcal{A}_z)$ (see Section \ref{sec:outc}); the first (DR-cc) uses correctly specified exposure and outcome model, the second (DR-cm) uses a wrong outcome model (see above), the third (DR-mm) a misspecified exposure model (i.e. $m_x(Z,C;\alpha)=\alpha_zZ+\alpha^T_cC$) and again the misspecified outcome model.
\end{description}
The results are displayed in Figure \ref{fig:sim1:bp} (top). They reveal, as predicted by Propositions 3 and 5, 
that Standard TSLS (TSLS) is robust against misspecification of the exposure and outcome model when $Z\cip C$, unlike the plug-in two-stage estimator (TS). As predicted in Section \ref{sec:exp}, the efficiency of Standard TSLS estimators can be improved by acknowledging that the exposure model is nonlinear (the relative efficiency of estimator LE-y-c versus TSLS is 0.38). However, the misspecified version (LE-y-m) is seriously biased.
The double-robust estimators (DR-cc, DR-cm and DM-mm), are somewhat less efficient as they avoid reliance on a correct outcome model, (e.g.\ the relative efficiency of estimator LE-cc versus LE-y-c is 1.14). This is even more pronounced under model misspecification, because semi-parametric efficiency is only guaranteed when the exposure and outcome models are correctly specified. However, by their double robustness, these estimators remain unbiased under such misspecification as they correctly exploit that $Z\cip C$.

Figure \ref{fig:sim1:bp} (bottom) shows results from a setting where the IV depends nonlinearly on covariates, and where the degree of model misspecification is more pronounced. In particular, data were generated and analysed as before, but with $P(Z=1|C,V)=\mbox{expit}(-1+V/2)$ and
$P(X=1|Z,U,V)=\Phi(Z+U+V-ZV+V^2/2)$, corresponding to a Nagelkerke pseudo-$R^2$ value of 0.13 and an F-value of 38 under the working model  $P(Z=1|C)=\mbox{expit}(\alpha_0+\alpha_1V)$. The findings are similar to before, but the nonlinear dependence of the instrument on covariates impairs the robustness of the Standard TSLS estimator, as predicted by Proposition 5 (its bias is 0.55).

\begin{figure}[htbp]
\centering
\includegraphics[width=\textwidth]{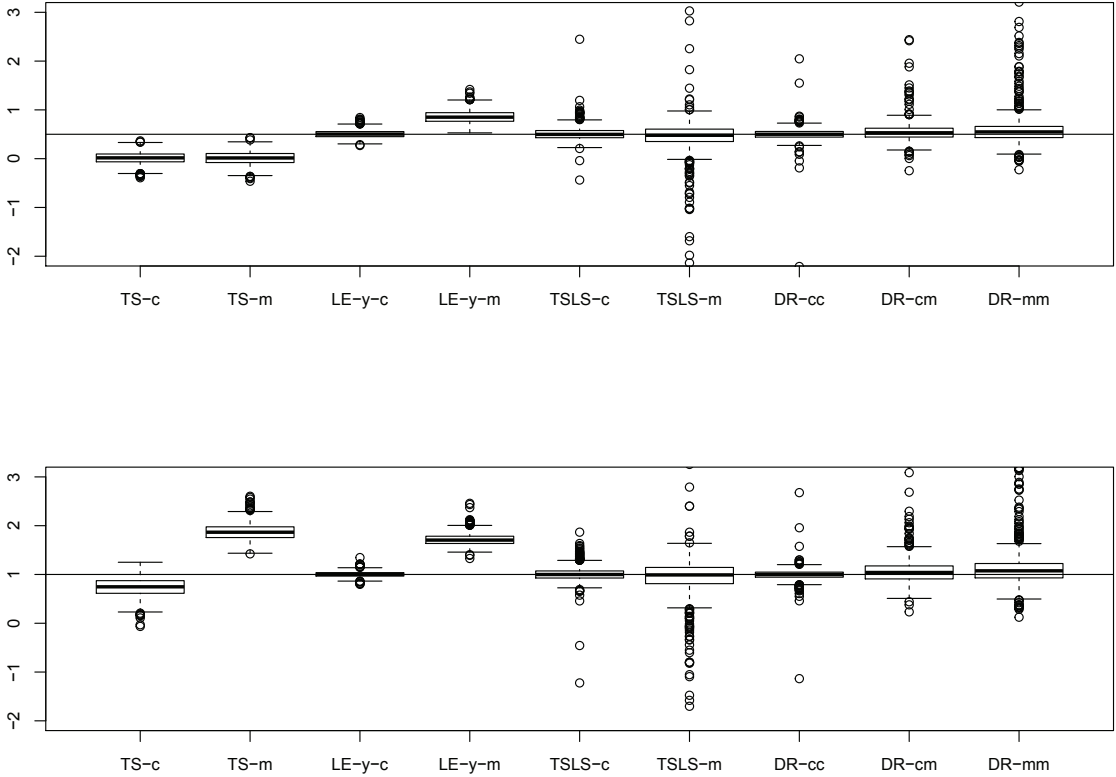}
\caption{Simulation results under effect modification. Boxplots of the main exposure effect estimators (top) and the exposure-covariate interaction estimators (bottom). Some of the outlying estimates are not visualised.}
\label{sim1:interactions}
\end{figure}

We also evaluated the behavior of the above estimators in the presence of effect modification. Data were generated as in the first simulation experiment above, but with $X$ being normal with mean $2Z+V+U-ZV+0.5V^2$ and $Y$ normal with mean $0.5X+XV-U-2V+V^2$. We evaluated the above estimators, but considered the Standard TSLS estimator with IVs $Z$ and $VZ$, and with $C=(1,V,V^2)$ (TSLS-c) and with $C=(1,V)$ (TSLS-m) to evaluate the impact of correct model specification. In addition, we considered the two-stage estimator with $C=(1,V,V^2)$ (TS-c) and with $C=(1,V)$ (TS-m) in the outcome model, but using a misspecified linear exposure model with main effects of $Z$ and $V$ only. The results are analogous as before. Figure \ref{sim1:interactions} (left) confirms the lack of robustness of two-stage estimators against misspecification of the exposure model, as opposed to Standard TSLS estimators (Figure \ref{sim1:interactions}, right). The best performance is seen for the locally efficient estimator under model $\mathcal{M}\cap\mathcal{A}_y$, but this estimator lacks robustness against misspecification of the outcome model. The Standard TSLS estimators and the double-robust estimators are much less efficient, especially under model misspecification. In the following sections, we will propose strategies to further improve performance.

\begin{center}
{\bf Figure \ref{fig:sim1:bp} and \ref{sim1:interactions} 
about  here.}
\end{center}

\section{Improved double-robust estimation}\label{sec:impro}

Consistency of the double-robust estimator of $\psi^*$ demands correct specification of either the outcome model $\mathcal{A}_y$ or the IV model $\mathcal{A}_z$; local efficiency demands correct specification of both these models, and additionally of models for the exposure distribution and conditional outcome variance. In practice all these models are typically somewhat misspecified. In Section \ref{sec:eem}, we therefore propose a strategy to guarantee efficiency within a subclass of double-robust estimators as soon as the IV model $\mathcal{A}_z$ is correctly specified. In Section \ref{subsec:brdr}, we propose strategies that aim to minimise locally  the bias of the double-robust estimator when both the outcome model $\mathcal{A}_y$ and the IV model $\mathcal{A}_z$ are misspecified. Throughout these sections, results are confined to the main effect structural model $\mathcal{M}^{const}$.

\subsection{Empirical efficiency maximisation}\label{sec:eem}

The semi-parametric efficient estimator of $\psi^*$, obtained by substituting the conditional expectations in (\ref{eeff}) by estimates under parametric models, is not guaranteed to outperform simpler CAN estimators (e.g. obtained by solving (\ref{eeg}) for $e(Z,C)=Z$ or by ignoring covariate information) under model misspecification, as we will see in the simulation study of Section \ref{sec:sim}.
Okui et al.\ (2012) proposed regression double-robust estimators that have no larger asymptotic variance than a given double-robust estimator, even under model misspecification. In this Section, we generalise their results with the potential for bigger efficiency gains in return. We will realise this by building on and extending the ideas behind empirical efficiency maximisation, a procedure originally proposed by Rubin and van der Laan (2008) and Cao, Tsiatis and Davidian (2009) in the missing data literature. In this subsection, we assume that model $\mathcal{A}_z$ is correctly specified. 

Let $\hat{\psi}(\alpha,\beta)$ be the double-robust estimator of $\psi^*$ obtained by solving estimating equation (\ref{eeg}) for a user-specified parameterisation 
 $e(Z,C;\alpha)$ of $e(Z,C)$, evaluated at the given values $\alpha$ (and $\beta$ indexing $m_y(C;\beta)$).
This may, but need not be guided by the form of the efficient index function given in (\ref{eeff}). 
For instance, for a scalar $Z$, one may postulate that $e(Z,C)$ is of the form $\alpha^T CZ$ for some $\alpha$. 
When the law of $Z$ given $C$ is known, then the asymptotic 
variance
of $\hat{\psi}(\alpha,\beta)$ under model $\mathcal{M}\cap\mathcal{A}_z$
equals
\begin{equation}
\label{effemp}
\frac{
\mbox{Var}\left(\left[e(Z,C;\alpha)
-E\left\{e(Z,C;\alpha)
|C\right\}\right]\left\{Y-m_y(C;\beta)
-\psi^*X\right\}\right)
}{
nE\left(\left[e(Z,C;\alpha)
-E\left\{e(Z,C;\alpha)
|C\right\}\right]X\right)^2
}.
\end{equation}
Let $\tilde{\alpha}$ and $\tilde{\beta}$ be the values of $\alpha$ and $\beta$, respectively, that minimise the empirical analog of (\ref{effemp}) with
$\psi^*$ substituted by a preliminary consistent estimator under model $\mathcal{M}^{const}\cap\mathcal{A}_z$, e.g. a G-estimator based on $e(Z,C)=Z$ and model $\mathcal{A}_y^{lin}$.
The proposition below then shows that $\hat{\psi}(\tilde{\alpha},\tilde{\beta})$ is a double-robust estimator which is at least as efficient as $\hat{\psi}(\alpha,\beta)$ for arbitrary $\alpha$ and $\beta$.
Key properties  that underlie the validity of the proposition are (a) that $\tilde{\beta}$ is CAN for $\beta^*$ under model $\mathcal{A}_y$; and (b) that $\hat{\psi}(\tilde{\alpha},\tilde{\beta})$ and $\hat{\psi}(\tilde{\alpha}^*,\tilde{\beta}^*)$  have the same asymptotic variance under model  $\mathcal{M}^{const}\cap\mathcal{A}_z$, with $\tilde{\alpha}^*$ and $\tilde{\beta}^*$ being the probability limits of $\tilde{\alpha}$ and $\tilde{\beta}$
(provided $\tilde{\alpha}$ and $\tilde{\beta}$ converge at faster than $n^{1/4}$ rate). 

\begin{prop} {\em Efficiency within a subclass of double-robust estimators}\\
\label{prop:rob.2sls3}
\setlength{\parindent}{0.3in} \setlength{\baselineskip}{24pt}
Let $\tilde{\alpha}$ and $\tilde{\beta}$ minimise the empirical version of (\ref{effemp}). 
Then the estimator $\hat{\psi}(\tilde{\alpha},\tilde{\beta})$ solving (\ref{eeg}) is CAN under model $\mathcal{M}^{const} \cap \left(\mathcal{A}_y \cup \mathcal{A}_z\right)$.

Moreover, when the law of $Z$ given $C$ is {\em known}, then we  have that for all $\alpha$ and $\beta$ 
\[ \lim_{n\rightarrow\infty} \mbox{Var}\left[\sqrt{n}\left\{\hat{\psi}(\tilde{\alpha},\tilde{\beta})-\psi^*\right\}\right]\leq \lim_{n\rightarrow\infty} \mbox{Var}\left[\sqrt{n}\left\{\hat{\psi}({\alpha},{\beta})-\psi^*\right\}\right].\]
\end{prop}
Proof: see Appendix B. $\square$

In Appendix B we further discuss the case where the law of $Z$ given $C$ is known only up to a finite-dimensional parameter.

Consider for instance the choices $e(Z,C;\alpha)=\alpha^T CZ$ and
$m_y(C;\beta)=\beta^T C$. Then by construction, $\hat{\psi}(\tilde{\alpha},\tilde{\beta})$ is at least as efficient as the estimator obtained by solving (\ref{eeg}) for the simple choices $e(Z,C)=Z$ and
$m_y(C)=0$, i.e.\ the estimator which ignores covariates. Hence, when $Z\cip C$, then the resulting approach will deliver a covariate adjustment strategy that is guaranteed to be at least as efficient as an unadjusted analysis. More generally, efficiency is - by construction - always attained within the subclass of estimators allowed by the models for $e(Z,C)$ and $m_y(C)$, but semi-parametric efficiency under model $\mathcal{M} \cap \left(\mathcal{A}_z \cup \mathcal{A}_y\right)$ is only attained when the efficient index function (\ref{eeff}) happens to equal $e(Z,C;\alpha)$ for some $\alpha$ and when $E(Y-\psi^*X|C)$ equals $m_y(C;\beta)$ for some $\beta$. 

Minimising the empirical analog of (\ref{effemp}) can generally be done numerically, but in special cases also by suitably modified regression techniques. 
For example, we show  in Appendix B of the Supplemental Materials that when $e(Z,C)=\alpha^{T}CZ$, then under certain assumptions minimising (\ref{effemp}) w.r.t. $\alpha$ can be done by fitting the regression model $E(X|Z,C)=\alpha^TC\left\{Z-E\left(Z|C\right)\right\}$ using ordinary least squares. 
Minimising (\ref{effemp}) w.r.t. $\beta$ is possible by fitting the regression model $E(Y-\psi^*X|C)=\beta^TC$ using weighted least squares  with weights $(\alpha^T C)^2\left\{Z-E\left(Z|C\right)\right\}^2$. 
%
%
The above procedure needs some modification when the law of $Z$ given $C$ is unknown and the model $\mathcal{A}_y$ is (possibly) misspecified.

The regression double-robust estimator of Okui et al. (2012) may be viewed as a special case of the above proposal. It fixes $\alpha$ at some given value (which may not minimise the asymptotic variance) and chooses $m_y(C;\beta)=\beta m_y(C)$ for some given $m_y(C)$. 

\subsection{Bias-reduced double-robust estimation}\label{subsec:brdr}

The efficiency results of Section \ref{sec:eem} are especially attractive when model $\mathcal{A}_z$ is known to hold, as is the case in certain study designs.
In other cases, bias becomes, arguably, a more dominant concern.  
Although there seems little hope that one can avoid bias in the estimation of $\psi^*$ when both working models $\mathcal{A}_z$ and $\mathcal{A}_y$ are misspecified, Vermeulen and Vansteelandt (2015) found that for quite a general class of double-robust estimators, surprisingly, the nuisance parameters indexing $\mathcal{A}_z$ and $\mathcal{A}_y$ can be estimated so as to target bias reduction. 
Briefly, they note that the asymptotic bias (Stefanski and Boos, 2002) of an estimator for $\psi^*$, evaluated at fixed nuisance parameters $\beta$ and $\gamma$, equals the expected value of its influence function $U(\psi^*,\beta,\gamma)$; for given $\alpha$, this is here:
\[
U(\psi,\beta,\gamma)=\frac{\left[e(Z,C;\alpha)
-E\left\{e(Z,C;\alpha)
|C;\gamma\right\}\right]\left\{Y-m_y(C;\beta)
-\psi X\right\}}{E\left(\left[e(Z,C;\alpha)
-E\left\{e(Z,C;\alpha)
|C;\gamma \right\}\right]X \right\}}.\]
Minimising the squared bias in the direction of $\beta$  thus amounts to setting the gradient
\[2E\left\{U(\psi^*,\beta,\gamma)\right\}E\left\{\frac{\partial U}{\partial \beta}(\psi^*,\beta,\gamma)\right\}\]
to zero. Although the first component cannot generally be made zero without knowing aspects of the data-generating law, interestingly, the second component delivers an unbiased estimating function for $\gamma$ (Vermeulen and Vansteelandt, 2015). This is so because, by the double-robustness, $U(\psi^*,\beta,\gamma)$ is mean zero for all $\beta$ at $\gamma^*$ when model $\mathcal{A}_z$ holds. The second component can thus be made zero empirically, by using it as a basis for estimation, as illustrated in the next paragraph. Under local misspecification of one of the working models (as formally defined in Vansteelandt et al. (2012)), this procedure reduces the order of the asymptotic bias. Under gross misspecification, it prevents inflation of the asymptotic bias, although one cannot exclude that other nuisance parameter estimators happen to deliver less biased effect estimators under some data-generating mechanisms. 

In the remainder of this section, we apply the bias-reduction procedure of Vermeulen and Vansteelandt (2015) to double-robust estimators in model $\mathcal{M}^{const}\cap(\mathcal{A}_z\cup\mathcal{A}_y)$.
For illustration, suppose that the instrument $Z$ is dichotomous with working model $P(Z=1|C;\gamma)=\mbox{expit}(\gamma^T C)$, that $m(C;\beta)=\beta^TC$, and let the index function $e(Z,C;\alpha)$ be of the form $Ze(C;\alpha)$ for some $e(C;\alpha)$ (as is the case for the efficient score for $\psi^*$ under model $\mathcal{M}^{const}\cap(\mathcal{A}_z\cup\mathcal{A}_y)$ when $E(X|Z,C)$ is linear in $Z$ and Var$(Y|Z,C)$ does not depend on $Z$). Taking the gradient of $U(\psi,\beta,\gamma)$ with respect to $\beta$ then results in estimating equations
\begin{eqnarray}
0&=&\sum_{i=1}^n \frac{\partial U_i(\gamma,\beta)}{\partial\beta}=
\sum_{i=1}^n e(C_i;\alpha)\left\{Z_i-P(Z_i=1|C_i;\gamma)\right\}C_i, \label{fegamma}
\end{eqnarray}
which are unbiased for $\gamma$. Since $\gamma$ and $\beta$ are of the same dimension, $\gamma$ can thus be estimated as the solution to this equation. Solving equation (\ref{fegamma}) ensures that 
\[
\sum_{i=1}^n e(C_i;\alpha)\left\{Z_i-P(Z_i=1|C_i;\gamma)\right\}m_y(C_i;\beta)=0
\]
so that the estimating equation for $\psi$ reduces to 
\begin{eqnarray*}
0&=&\sum_{i=1}^n e(C_i;\alpha)\left\{Z_i-P(Z_i=1|C_i;\gamma)\right\}\left\{Y_i-m_y(C_i;\beta)-m_{}(C_i;\psi^*)X_i\right\}\\
&=&\sum_{i=1}^n e(C_i;\alpha)\left\{Z_i-P(Z_i=1|C_i;\gamma)\right\}\left\{Y_i-m_{}(C_i;\psi^*)X_i\right\}
\end{eqnarray*}
which no longer involves $\beta$. Bias-reduced estimation of $\gamma$ then overcomes the need to estimate $\beta$, and thereby prevents that the choice of estimator of $\beta$ amplifies bias. 
Solving (\ref{fegamma}) may not be straightforward for certain data sets. 
We therefore adapt the proposal of Vermeulen and Vansteelandt (2015) by extending the logistic regression model for $Z$ to
\[P(Z=1|C;\gamma)=\mbox{expit}\left\{\gamma^T C+\theta^TCe(C;\alpha)\right\}.\]
This model contains the original working model $\mathcal{A}_z$ (corresponding to $\theta=0$). Moreover, fitting this model using the default maximum likelihood procedure 
has the effect of making the identity (\ref{fegamma}) hold, as the latter corresponds with the score for the coefficient of $e(C;\alpha)C$. 
The resulting procedure will be referred to as BR-$\gamma$. 

When using the procedure BR-$\gamma$, we continue to estimate $\alpha$ indexing $e(C;\alpha)$ as explained in Section \ref{sec:eem}.
Although now, we no longer assume that model $\mathcal{A}_z$ is correctly specified, estimating $\alpha$ in this manner still has the effect of minimising the asymptotic variance of the double-robust estimator across all values of $\alpha$. This is because the procedure BR-$\gamma$ sets the gradient of the influence function w.r.t. $\beta$ equal to zero, so that there is no need to account for the estimation of $\beta$ in the calculation of the asymptotic variance (Vermeulen and Vansteelandt, 2015). 
Because BR-$\gamma$ moreover employs maximum likelihood estimation under a particular extended model for the IV, conservative standard errors can be obtained as 1 over root $n$ times the empirical standard deviation of $U(\gamma,\beta)$, ignoring the estimation of $\gamma$ (Rotnitzky, Li and Li, 2010). Alternatively, robust sandwich standard errors can be calculated, or the bootstrap can be used.

We also considered a related approach whereby  we estimated $\gamma$ using maximum likelihood and $\beta$ by setting the gradient of the influence function $U(\gamma,\beta)$ with respect to $\gamma$ to zero. This results in the following unbiased estimating equations for $\beta$:
\begin{eqnarray}
0&=&\sum_{i=1}^n \frac{\partial U_i(\psi^*,\gamma,\beta)}{\partial\gamma}=\sum_{i=1}^n e(C_i;\alpha)\left\{Y_i-m_y(C_i;\beta)
-\psi^*X_i\right\}\Gamma_i
\label{febeta}
\end{eqnarray}
for
\begin{eqnarray*}
\Gamma_i &=& \left\{Z_i-P(Z_i=1|C_i;\gamma)\right\}E\left[e(C_i;\alpha)P(Z_i=1|C_i;\gamma)P(Z_i=0|C_i;\gamma)C_i X_i\right]\\
&&-P(Z_i=1|C_i;\gamma)P(Z_i=0|C_i;\gamma)C_iE\left[e(C_i;\alpha)\left\{Z_i-P(Z_i=1|C_i;\gamma)\right\}X_i\right]
\end{eqnarray*}
It can be verified that the effect of the factor $\Gamma_i$ is to eliminate $\psi^*$ from the estimating equation so that knowledge of the truth $\psi^*$ is not needed for estimating $\beta$.
This approach is designed to locally minimise the bias of the double-robust estimator in the direction of $\gamma$, at the maximum likelihood estimate $\hat{\gamma}$.
To solve (\ref{febeta}), we jointly fit 
an extended linear model for the outcome $Y-\psi^* X$ with covariates $C$ and $e(C;\alpha)P(Z=1|C;\gamma)P(Z=0|C;\gamma)C$ using ordinary least squares (where, again, the choice of $\psi^*$ does not affect results), and  the (double-robust) estimating equation for $\psi$. This has the effect of making the identity (\ref{febeta}) hold. 
The resulting procedure will be referred to as BR-$\beta$. By setting the gradient of the influence function $U(\gamma,\beta)$ with respect to $\gamma$ equal to zero, it need not adjust for the estimation of $\gamma$ (Vermeulen and Vansteelandt, 2015). However, the uncertainty in the estimate of $\beta$ must be acknowledged using sandwich standard errors or the bootstrap.

\section{Simulation study}\label{sec:sim}


We conducted a simulation experiment with $n=500$ independent measurements on mutually independent and standard normal covariates $U$ and $V$, $Z$ dichotomous with $P(Z=1|V)=\mbox{expit}(-1+V/2+\lambda_z V^2/3)$, $X$ normal with mean $Z+U+V-ZV+\lambda_x V^2$ and $Y$ normal with mean $X-U-V+\lambda_y V^2$. Assuming a linear IV model with $m_{}(C;\psi)=\psi$, we then evaluated the following estimators:
\begin{enumerate}
\item TSLS: the Standard TSLS estimator using $(Z,VZ)^T$ as IV vector,  based on a linear model for the exposure, involving main effects of $V,Z$ and their interaction, and a linear model for the outcome involving main effects of $V$ and the fitted value from the first stage regression. Including the $VZ$ interaction in the first stage model ensures a fairer comparison with the subsequent estimators so that for all estimators misspecification of the exposure model is only due to omitting $V^2$.
\item Loc Eff: the locally efficient double-robust estimator (assuming homoscedasticity) based on a logistic model for $Z$ with a main effect of $V$, a linear model for $X$ with a main effect of $Z$ and $V$ and their interaction, and a linear outcome model (i.e., $m_y(C)=\beta^TC$ with $C=(1,V)^T$), all fitted using maximum likelihood.
\item Emp Eff: the locally efficient double-robust estimator using the same fitted model for $Z$ as before, but using working models $e(Z,C)=\alpha^TCZ$ and
$m_y(C)=\beta^TC$ fitted using empirical efficiency maximization (ignoring estimation of the model for $Z$, which is suboptimal when the outcome model is misspecified).
\item BR-$\beta$, BR-$\gamma$: the double-robust estimator with $\alpha^*$ estimated using empirical efficiency maximization, but with either the outcome model or the IV model fitted using bias-reduced estimation.
\end{enumerate}
To obtain the estimators Loc Eff and Emp Eff, the TSLS estimator was used as a starting value; the obtained estimate was then updated a single time. In the calculation of BR-$\beta$, BR-$\gamma$ was used as a starting value as it was easy to obtain and generally performing well.

Table \ref{tab:sim2} shows the simulation results based on 1000 simulations. When all working models are correctly specified (i.e. $\lambda_z=\lambda_x=\lambda_y=0$), then all estimators have nearly identical performance to Standard TSLS. This is theoretically expected, because they are based on correctly specified working models in the calculation of the efficient score and are therefore asymptotically equivalent. 
When only the outcome model is misspecified (i.e. $\lambda_z=\lambda_x=0,\lambda_y\ne 0$), then the TSLS estimator is  biased (as the instrument distribution does not satisfy Proposition \ref{prop:rob.2sls2}) with larger standard deviation than the double-robust estimators, which were all unbiased. Bias-reduced estimation of the outcome model (BR-$\beta$) resulted in major efficiency gains in this case. 
When only the exposure model was misspecified (i.e. $\lambda_z=\lambda_y=0,\lambda_x\ne 0$), then as theoretically predicted, the TSLS estimator and the double-robust estimators  continue to be unbiased, but the performance of the locally efficient double-robust estimator was sometimes very poor because its efficiency is only attained at a correctly specified model for the exposure. In this case, drastic improvements were obtained via empirical efficiency maximization, because this strategy guarantees efficiency within a subclass of estimators, regardless of correct specification of an exposure model. The efficiency of  the resulting double-robust estimator was sometimes better, sometimes worse than that of TSLS.
When only the IV model was misspecified (i.e. $\lambda_x=\lambda_y=0,\lambda_z\ne 0$), then all estimators were unbiased because of the double-robustness of the estimators and the fact that the TSLS estimator does not rely on correct specification of an IV model; all estimators had nearly identical performance in this case. When both the exposure and outcome model are misspecified (i.e. $\lambda_z=0,\lambda_x\ne 0,\lambda_y\ne 0$), then again TSLS is  biased, unlike the double-robust estimators. The locally efficient estimator behaved poorly in this case and is greatly outperformed by empirical efficiency maximisation, which again performs best in combination with bias-reduced estimation of the outcome model. When all models were misspecified, then also the double-robust estimators were subject to bias. However, bias-reduced estimation of either the outcome model or the IV model resulted in bias reductions and efficiency gains. This is not surprising for BR-$\gamma$ because the extended IV model happened to contain the truth: indeed, the inclusion of the covariate $e(C;\alpha)^TC$ in the instrument model was tantamount to the inclusion of $V^2$. For BR-$\beta$, where the extended outcome model did not contain the truth, this confirms that the proposed procedure reduces bias under model misspecification.

To further evaluate the bias-reduced estimation strategy, we additionally ran simulations under extreme misspecifications, such that both extended outcome and IV models did not contain the truth. In particular, we generated $n=500$ independent measurements on mutually independent and standard normal covariates $U$ and $V$, $Z$ dichotomous with $P(Z=1|V)=1-\exp\left\{-\exp(-1+V/2-V^2/2+\lambda_z V^2/8)\right\}$, $X$ normal with mean $Z+U+V-ZV+2V^2+2ZV^2+2\lambda_x V^3$ and $Y$ normal with mean $X-U-V-2V^2+2\lambda_y V^3$. 

\begin{figure}[htbp]
\centering
\includegraphics[width=0.9\textwidth]{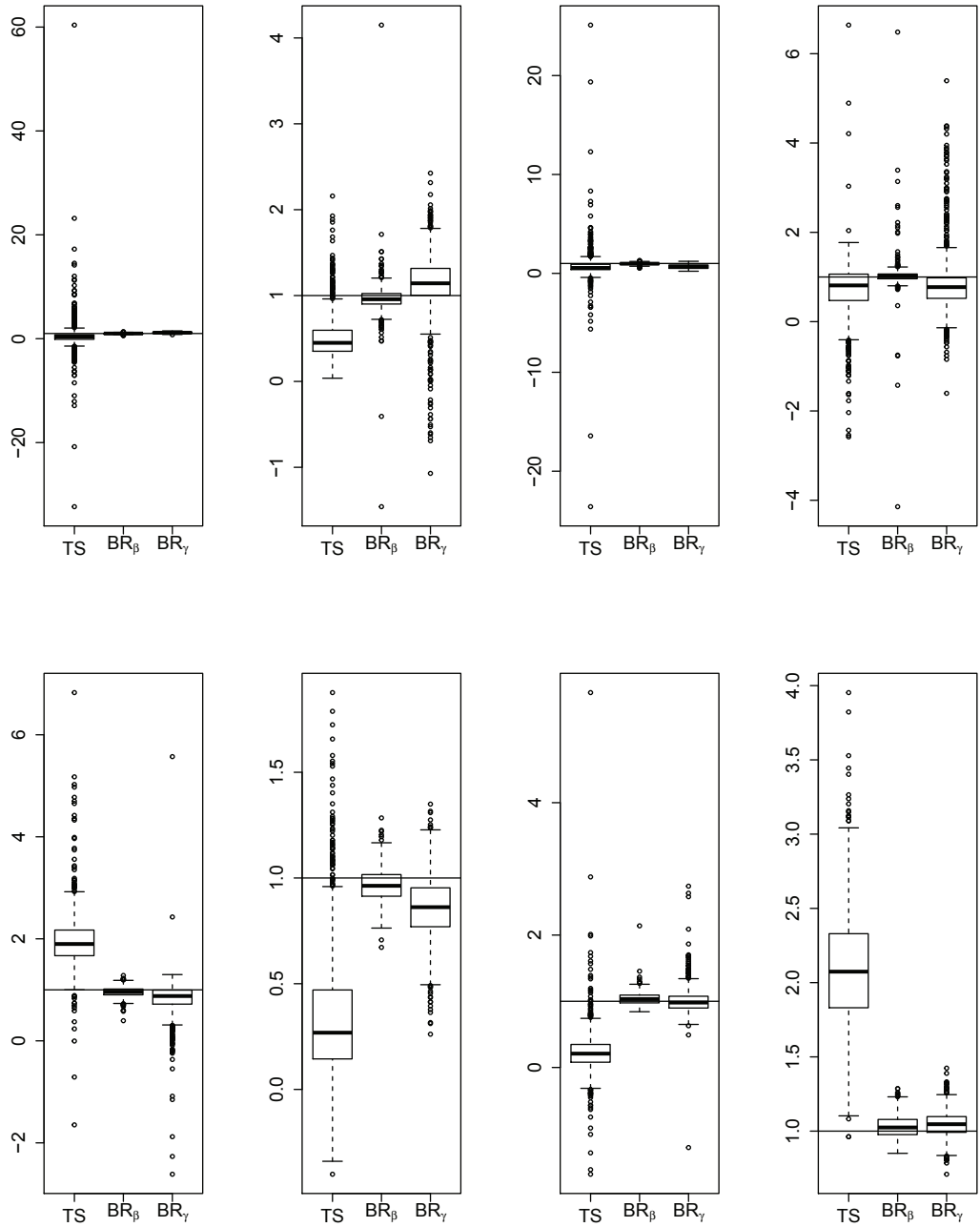}
\caption{Boxplots of the two-stage estimator (TS), the double-robust estimator based on empirical efficiency maximization and bias-reduced nuisance parameter estimators (BR) under extreme model misspecification. }
\label{SimIVCov3}
\end{figure}

The working models were the same as before. The results are visualised in Figure \ref{SimIVCov3} for all combinations of $\lambda_x,\lambda_y$ and $\lambda_z$ in $\{-1,1\}$, and confirm the previous findings. The locally efficient double-robust estimator had very poor performance and, while empirical efficiency maximization resulted in major efficiency gains, it was still much worse than TSLS estimation. 
For instance, in the setting of Figure \ref{SimIVCov3} (top, left), the locally efficient double-robust estimator had bias and standard deviation of -32.7 and 450, as opposed to -0.61 and 4.1 with empirical efficiency maximization, and -0.54 and 3.1 with TSLS.
In combination with bias-reduced estimation, most of the bias disappeared and variance was often greatly reduced (see Figure \ref{SimIVCov3}).

\vspace*{-.25cm}
\begin{center}
{\bf Table \ref{tab:sim2} and Figure \ref{SimIVCov3} about here.}
\end{center}

\section{Illustration}

We illustrate the proposed methodology on a sample of 3010 working men aged between 24 and 34 who were part of the 1976 wave of the US National Longitudinal Survey of Young Men (Card, 1995). In particular, we will estimate the effect of years of education on the log of hourly wages in 1976 ($Y$). Following Card (1995), we use as an IV an indicator if the individual lived close to a college that offered 4 year courses in 1966 ($Z$). All reported analyses are adjusted for covariates ($C$) years of labour market experience and its square, marital status, an indicator if the individual is black, as well as various measures of geographical location in 1966 and 1976. Twelve years of education was most common (33\%) in this study and was therefore used as a reference class by defining $X$ to be the difference between the years of education and 12. 

The log of hourly wages is reasonably normally distributed with mean 6.3 (SD 0.44), and is on average 0.075 (95\% CI 0.068 to 0.082) higher per extra year of education, after linear regression adjustment for years of labour market experience, marital status, race and geographical location in 1966 and 1976. The partial correlation between education and the IV is 0.066. Below, we will report the results from IV analysis with 95\% percentile-based confidence intervals based on the nonparametric bootstrap with 1000 resamples.

Standard TSLS analysis yields an education effect of 0.13 (SE 0.067, 95\% CI 0.029 to 0.28) on the average log of the hourly wage, corresponding with a one-year increase in education. Because the instrument is dichotomous and strongly associated with covariates, its expectation is likely nonlinear  in the covariates. The Standard TSLS estimator is therefore sensitive to correct specification of the role of covariates in the outcome model. We thus evaluate the double-robust estimators based on a logistic regression model for the IV. The locally efficient G-estimator equals 0.10 (SE 0.044, 95\% CI  0.025 to 0.18). Like the double-robust estimator based on empirical efficiency maximization (0.088, SE 0.045, 95\% CI 0.0063 to 0.18), it is much more efficient than the Standard TSLS estimator. Further, more minor efficiency gains are obtained through the proposed bias reduction strategies. In particular, we find that BR-$\gamma$ equals 0.092 (SE 0.041, 95\% CI 0.010 to 0.18) and BR-$\beta$ equals 0.095 (SE 0.043, 95\% CI 0.0063 to 0.19). 
      
\section{Discussion}

In this article, we have argued that Standard TSLS estimation, unlike many variations of the two-stage approach to estimation with an IV, is often robust against misspecification of the working models for the exposure and outcome.  However, this robustness may come at the expense of a loss of precision, which can be considerable when, for instance, the exposure mean is nonlinear in the instrument and/or covariates, e.g.\ when the exposure is binary, multinomial or count data.  
Moreover, the suggested robustness of the Standard TSLS estimators is limited to specific data-generating mechanisms: 
robustness against misspecification of the outcome model is for instance lost in Standard TSLS estimators when the IV is nonlinear in covariates.  
We also demonstrated that another strength of Standard TSLS, not generally shared by other two-stage estimators, is that including covariates will asymptotically not reduce, and typically improve, efficiency when instrument and covariates are known to be independent and in the absence of effect modification.

In contrast, locally efficient double-robust IV estimators confer robustness against model misspecification in a wider class of data generating mechanisms.  For instance, an attractive alternative, when instruments and covariates are known to be independent, is the estimator obtained by empirical efficiency maximisation: it is guaranteed consistent and efficient relative to a subclass of all CAN estimators.  
In other situations one should arguably worry more about bias than efficiency. We have shown that major improvements can be achieved by combining empirical efficiency maximisation with bias-reduced double-robust estimation. The resulting estimators have a very stable performance with considerable robustness against misspecification of all models for the instrument, exposure and outcome; their standard errors can be computed relatively easily using sandwich estimators. We are hopeful that by extending these results to double-robust estimators in nonlinear IV models (Robins, 1994; Vansteelandt et al., 2010), we will be able to improve the performance of IV estimators in these more complex settings where difficulties of estimation are common (Vansteelandt et al., 2011; Burgess et al., 2014). 

There are some limitations to our work. Our results are asymptotic and do not take into account the problem of `weak instrument / small sample' bias (Bound et al., 1995). This may in practice exacerbate the problem of bias due to model misspecification. There are a number of variations on two-stage estimators that are designed to address this problem, such as e.g.\  limited information maximum likelihood (Anderson, 2004), but these will not generally exhibit comparable robustness towards model misspecification. It would be an important area for future research to tackle both sources of bias simultaneously. Related to this, although the results on empirical efficiency maximisation appear to suggest that it is beneficial to adjust for all available covariates $C$ when $Z\cip C$, the performance of the resulting estimators my be affected in the presence of high-dimensional covariates. Whether and how to best select covariates in such cases, as well as in settings where it is not known whether $Z\cip C$, constitutes an important area for future research.

\section*{References}

\begin{description}
\item
Anderson, T.W.\ (2005).
Origins of the limited information maximum likelihood and two-stage least squares estimators. \textit{Journal of Econometrics}, \textbf{127}, 1-16.
\vspace*{-.25cm}
\item
Angrist,  J.D.,   Imbens, G.W.,  and     Rubin, D.B. (1996). Identification of causal effects using instrumental variables. {\it Journal of the American Statistical Association}, {\bf 91}, 
 444-455.
\vspace*{-.25cm} \item
Bound,  J.,  Jaeger,  D.A.  and Baker,  R.M.\ (1995).  Problems  with  instrumental   variables   estimation   when   the   correlation   between  the  instruments  and  the  endogenous  variable  is weak. {\it Journal of the American Statistical Association}, {\bf 90}, 443-50.
\vspace*{-.25cm} \item
 Bowden, J.\ and Vansteelandt, S. (2011). Mendelian randomisation analysis of case-control data using Structural Mean Models. \textit{Statistics in Medicine}, {\bf 30}, 678-694.
\vspace*{-1cm} \item
 Bowden, R.J.\ and Turkington, D.A.\ (1985). {\it Instrumental Variables}. Cambridge University Press.
\vspace*{-.25cm} \item
 Burgess, S., Granell, R., Palmer, T.M., Sterne, J.A.C. and Didelez, V. (2014).
Lack of identification in semi-parametric instrumental variable models with binary outcomes. \textit{American Journal of Epidemiology}, 
\vspace*{-.25cm} \item
 Card, D. (1995). \textit{Using geographic variation in college proximity to estimate the return to schooling.} In L. Christophides, E. Grant and R. Swidinsky (eds.), Aspects of Labour Market Behaviour: Essays in Honour of John Vanderkamp.
\vspace*{-.25cm} \item
 Cao, W.H., Tsiatis, A.A. and Davidian, M. (2009). Improving efficiency
and robustness of the double-robust estimator for a population mean with
incomplete data. \textit{Biometrika}, \textbf{96}, 723-734.
\vspace*{-.25cm} \item
 Chamberlain, G. (1987). Asymptotic efficiency in estimation with conditional moment restrictions. \textit{Journal of Econometrics}, {\bf 34}, 305-334.
\vspace*{-.25cm} \item
 Clarke, P.S. and Windmeijer, F. (2010).
Identification of causal effects on binary outcomes using structural mean models. \textit{Biostatistics}, {\bf  11}, 756-770.
\vspace*{-.25cm} \item
 Didelez, V. and Sheehan, N.A. (2007). {M}endelian randomization as an instrumental variable approach to  causal inference. {\it Statistical Methods in Medical Research,} 16, 309-330. 
\vspace*{-.25cm} \item
 Didelez, V., Meng, S. and Sheehan, N.A. (2010). Assumptions of IV methods for observational epidemiology. {\it Statistical Science}, \textbf{25}, 22-40. 
\vspace*{-.25cm} \item
Fischer-Lapp, K., and Goetghebeur, E.\ (1999). Practical properties of some structural mean analyses of the effect of compliance in
randomized trials. {\it Controlled Clinical Trials}, {\bf 20}, 531-546.
\vspace*{-.25cm} \item
Greenland, S. (2000). An introduction to instrumental variables for epidemiologists. {\it International Journal of Epidemiology,}  29 (4), 722-729.
\vspace*{-.25cm} \item
 Hern\'an, M.A. and Robins, J.M. (2006). Instruments for causal inference - An epidemiologist's dream? {\it Epidemiology}, {\bf  17}, 360-372.
\vspace*{-.25cm} \item
Kleibergen, F. and Zivot, E.\ (2003). Bayesian and classical approaches to instrumental variable regression.
{\it Journal of Econometrics}. {\bf 114}, 29-72.
\vspace*{-.25cm} \item
Mullahy,  J.  (1997).  Instrumental  variable  estimation  of count data models: Application to models of cigarette smoking behaviour. {\it Review of Economics and Statistics,} 79, 586-593.
\vspace*{-.25cm} \item
 Okui, R., Small, D.S., Tan, Z.Q. and Robins, J.M. (2012). Doubly robust instrumental variable regression. \textit{Statistica Sinica}, {\bf 22}, 173-205.
\vspace*{-.25cm} \item
 Pearl, J. (2009). {\em Causality: Models, Reasoning and Inference.} 
Cambridge University Press, New York, NY, USA, 2nd edition.
\vspace*{-.25cm} \item
 Robins, J.M. (1994). Correcting for non-nompliance in randomized trials using
structural nested mean models. {\it Communications in Statistics: Theory and Methods} {\bf 23}, 2379-2412.
\vspace*{-.25cm} \item
Robins JM. (2000). Robust estimation in sequentially ignorable missing data and causal inference models. \textit{Proceedings of the American Statistical Association}, Section on Bayesian Statistical Science 1999, pp. 6-10.
\vspace*{-.25cm} \item
Robins, J.M., and Rotnitzky, A.\ (2001). Comment on ``Inference for semiparametric models: Some questions and an answer,'' by P.J. Bickel and J. Kwon. 
{\it Statistica Sinica}, {\bf 11}, 920-936.
\vspace*{-.25cm} \item
 Rotnitzky, A., Li, L.L., and Li, X.C. (2010). A note on overadjustment in inverse probability weighted estimation. \textit{Biometrika}, \textbf{97}, 997-1001.
\vspace*{-.25cm} \item
 Rubin, D.B. and van der Laan, M.J. (2008). Empirical efficiency
maximization: improved locally efficient covariate adjustment in randomized
experiments and survival analysis. \textit{International Journal of
Biostatistics}, \textbf{4}, 1-40.
\vspace*{-.25cm} \item
 Stefanski, L.A. and Boos, D.D. (2002). The calculus of M-estimation.
\textit{The American Statistician}, \textbf{56}, 29-38.
\vspace*{-.25cm} \item
 Tchetgen Tchetgen, E.J., Walter, S., Vansteelandt, S., Martinussen, T. and Glymour, M. (2015). Instrumental variable estimation in a survival context. \textit{Epidemiology}, \textbf{26}, 402-410. 
\vspace*{-.25cm} \item
 Vansteelandt, S., Bowden, J., Babanezhad, M. and Goetghebeur, E. (2011). On instrumental variables estimation of causal odds ratios. \textit{Statistical Science}, \textbf{26}, 403-422.
\vspace*{-1cm} \item Vansteelandt, S., Bekaert, M. and Claeskens, G. (2012). On model selection and model misspecification in causal inference. \textit{Statistical Methods in Medical Research}, {\bf 21}, 7-30.
\vspace*{-.25cm} \item
 Vermeulen, K. and Vansteelandt, S. (2015). Bias-reduced doubly robust estimation. \textit{Journal of the American Statistical Association}, in press.
\vspace*{-.25cm} \item
 Wooldridge, J. (2002). \textit{Econometric Analysis of Cross Section and Panel Data}, MIT Press, Cambridge, MA.
\end{description}

\newpage 

{\scriptsize
\begin{table}[htbp]
\centering
\caption{Empirical bias and standard deviation of the two-stage estimator (TS), the locally efficient double-robust estimator (Loc Eff), the double-robust estimator based on empirical efficiency maximization (EEM) and these same estimators that employ bias-reduced nuisance parameter estimators (BR). The superscript number between brackets refers to the number of severely outlying estimates that were eliminated in the calculation of bias and empirical standard deviation.}
\label{tab:sim2}
\begin{tabular}{lrrrrrrrr}\hline
 & $\lambda_x$ & $\lambda_y$ & $\lambda_z$ & TS & Loc Eff & EEM & BR-$\beta$ & BR-$\gamma$ \\
\hline
Bias & 0 & 0 & 0 & 0.0033 & 0.0043 & 0.0044& 0.0041  & 0.0042\\
& 0 & 1 & 0& -0.55 & -0.0092 & -0.035 & 0.0024 & -0.017\\
& 0 & -1 & 0 & 0.56 & 0.018 & 0.044 & 0.0059 & 0.026\\
& 1 & 0 & 0 & 0.000073 & 0.013 & 0.0058 & 0.0037 & 0.0046\\ 
& -1 & 0 &0 & 0.0013 &  0.0074 &  0.0043 &   0.0048 &  0.0044\\
& 0 & 0 & 1 & -0.00057 & -0.00033 & $9.0 \ 10^{-5}$ & -$9.1 \ 10^{-5}$ &  0.00029 \\
& 0 & 0 & -1 &  0.0053 & 0.0055 & 0.0095 & 0.0050 & 0.0045 \\
& 1 & 1 & 0 & 0.15 & 0.11$^{(2)}$ & -0.040 & 0.0039 & -0.019\\ %
& -1 & 1 & 0 & -0.41 & 0.012 & -0.021 &  0.0016 &  -0.013 \\
& 1 & -1 & 0 & -0.15 &  -0.095$^{(4)}$ &  0.051 &  0.0038 & 0.028\\ %
& -1 & -1 & 0 &  0.41 & 0.0030 & 0.030 & 0.0079 &  0.022\\
& 1 & 1& 1 & 0.34 & 0.36 &0.11 & 0.021 &-0.00028 \\
& -1 & 1 & 1 & -0.35 & -14$^{(2)}$ & -0.40&   0.024&  0.00073 \\
& 1 & -1 & 1 & -0.34 & -0.36 & -0.11 & -0.023 & -0.00059 \\
& -1 & -1 & 1 & 0.35 &15$^{(2)}$ & 0.86 & -0.023 & 0.0015 \\
& 1 & 1 & -1 & -0.94 & 0.59$^{(1)}$ & -0.48 & 0.019 & 0.0057\\ %
& -1 & 1 & -1 & -0.36 & -0.084 & -0.085 & 0.018 & 0.0048\\
& 1 & -1 & -1 & 0.94 & -0.20$^{(2)}$  & 0.50 & -0.0081 & 0.0039 \\ %
& -1 & -1 & -1 & 0.37 & 0.10 & 0.10 & -0.0086 & 0.0036\\
\hline
SD & 0 & 0 & 0 &  0.11 & 0.11 &  0.11 & 0.11 & 0.11\\
& 0 & 1 & 0& 0.24 &  0.18 & 0.17 & 0.12 & 0.17\\
& 0 & -1 & 0 & 0.28 & 0.19 & 0.19 & 0.12 & 0.18\\
& 1 & 0 & 0 & 0.18 & 0.82$^{(4)}$ & 0.12 & 0.12 & 0.12 \\
& -1 & 0 &0 & 0.068 &   0.12 & 0.11 & 0.11 & 0.11\\
& 0 & 0 & 1 & 0.094 & 0.097 &  0.11 & 0.097 &  0.097\\
& 0 & 0 & -1 &  0.13 &  0.13 & 0.14 & 0.13 &   0.13\\
& 1 & 1 & 0 & 0.46 & $1.9^{(2)}$ &  0.19 & 0.12 & 0.17\\
& -1 & 1 & 0 & 0.11 & 0.23 & 0.17 & 0.12 & 0.17 \\
& 1 & -1 & 0 & 0.48 & 1.2$^{(4)}$ & 0.20 & 0.12 & 0.18\\
& -1 & -1 & 0 & 0.14 & 0.22 & 0.17 & 0.12 & 0.17\\
& 1 & 1& 1 & 0.15 & 0.12 &0.16 & 0.10& 0.14\\
& -1 & 1 & 1 & 0.15 & 120$^{(2)}$ & 8.30&  0.11& 0.14\\
& 1 & -1 & 1 & 0.14 & 0.10 & 0.16 &  0.099&  0.13\\
& -1 & -1 & 1 & 0.16 & 140$^{(2)}$ & 12&  0.10&  0.13\\
& 1 & 1 & -1 & 1.3 & 11$^{(1)}$ & 0.85 & 0.13 & 0.17\\
& -1 & 1 & -1 & 0.098 & 0.17 & 0.18 & 0.14 & 0.17\\
& 1 & -1 & -1 & 1.5&  5.6$^{(2)}$  & 0.98 &  0.13 &  0.17\\
& -1 & -1 & -1 & 0.12 & 0.17 & 0.18 & 0.13 & 0.16\\
\hline
\end{tabular}
\end{table}}

\end{document}


\author{Stijn Vansteelandt \\[1ex] 
\textit{Department of Applied Mathematics, Computer Sciences and Statistics} \textit{\ }%
\\
\textit{Ghent University, Belgium} \and 
and Vanessa Didelez \\
\textit{School of Mathematics} \textit{\ }%
\\
\textit{University of Bristol, U.K.} 
}
\title{
Supplemental Material for \\ `Robustness and efficiency of covariate adjusted linear instrumental variable estimators' }
\date{}

\maketitle

\bigskip \setlength{\parindent}{0.3in} \setlength{\baselineskip}{24pt}

\subsection*{Appendix A: Robustness and Efficiency of Two-Stage Estimation}

{\em Observational equivalence of IV models in Section 2:} We show that models (1) and (3) are observationally equivalent under a mild weakening of the IV assumptions that replaces independence by mean independence. 
It follows from the Appendix of Robins and Rotnitzky (2004) that model (3) along with the IV assumptions is the same model for the observed data law as the model defined by $E\left\{Y-m(C;\psi^*)X|Z,C\right\}=E\left\{Y-m(C;\psi^*)X|C\right\}$. That model (1) implies this restriction is immediate. To show the opposite, we must show that for each observed data law satisfying this restriction, there is a law of $(Y,X,U,Z,C)$ that satisfies (1) and the IV-assumptions, and which marginalises to the observed data law. This is immediate upon choosing $\omega(C,U)=E\left\{Y-m(C;\psi^*)X|C\right\}+U$ with $U=Y-m(C;\psi^*)X-E\left\{Y-m(C;\psi^*)X|C\right\}$. 
Indeed, it is immediate from the expression for $U$ that $E(U|Z,C)$ equals zero and hence does not depend on $Z$.

\smallskip

{\em Proof of Proposition 2:}
We start by deriving the semi-parametric efficient estimator that jointly fits the first and second stage models; then we show that under the given assumptions Standard TSLS is such an estimator. 

Under a mild weakening of the IV assumptions that replaces independence by mean independence, it follows from the previous paragraph that model $\mathcal{M}\cap\mathcal{A}_x\cap\mathcal{A}_y$ is the same model for the observed data as the standard (multivariate) conditional mean model defined by $\mathcal{A}_x$ and $E\left\{Y-m(C;\psi^*)X|Z,C\right\}=E\left\{Y-m(C;\psi^*)X|C\right\}=m_y(C;\beta^*)$, where the latter identity is equivalent to (10). Hence, it follows from general results on conditional mean models (Chamberlain, 1987) that 
all CAN estimators of $(\alpha^*,\beta^*,\psi^*)$ in model $\mathcal{M}\cap\mathcal{A}_x\cap\mathcal{A}_y$ are asymptotically equivalent to 
the solution to an estimating equation of the form
\begin{equation}
\label{eelin}
0=\sum_{i=1}^n d(Z_i,C_i)\left(\begin{array}{c} X_i-m_x(Z_i,C_i;\alpha) \\ Y_i-m_y(C_i;\beta)-m_{}(C_i;\psi)m_x(Z_i,C_i;\alpha) \end{array}\right),
\end{equation}
for some choice of $d(Z_i,C_i)$, an arbitrary function of the dimension of $(\alpha^*,\beta^*,\psi^*)$. 
This equation may equivalently be written as 
\begin{eqnarray*}
0&=&\sum_{i=1}^n  \left[
d_x(Z_i,C_i)\left\{X_i-m_x(Z_i,C_i;\alpha) \right\}  \right. \nonumber
\\&&+
 \left.
 d_y(Z_i,C_i)\left\{Y_i-m_y(C_i;\beta)-m_{}(C_i;\psi)m_x(Z_i,C_i;\alpha)\right\}  \right],
\end{eqnarray*}
for conformable functions $d_x(Z_i,C_i)$ and $d_y(Z_i,C_i)$.
The semi-parametric efficient estimator  of $(\alpha^*,\beta^*,\psi^*)$ in this class 
is obtained by choosing $d(Z,C)$ in the above expression (\ref{eelin}) equal to  (Chamberlain, 1987)
\[
d_{\rm opt}(Z,C)=\left(\begin{array}{cc} \partial m_x(Z,C;\alpha^*)/\partial\alpha & m_{}(C;\psi^*)\partial m_x(Z,C;\alpha^*)/\partial\alpha \\ 0 & \partial m_y(C;\beta^*)/\partial\beta 
\\ 0 & m_x(Z,C;\alpha^*)\partial m_{}(C;\psi^*)/\partial\psi\end{array}\right)\mbox{Var}\left\{\left(\begin{array}{c} X\\Y\end{array}\right) \mid Z,C\right\}^{-1}.
\]
Two-stage estimators for $(\alpha^*,\beta^*,\psi^*)$ solve the equations (11) and (12)
for index functions $e_x(Z,C)$ of the dimension of $\alpha$ and $e_y(Z,C)$ of the dimension of $(\beta,\psi)$. 
As this does not necessarily conform with the structure of $d_{\rm opt}(Z,C)$ above, the semi-parametric efficient estimator can lie outside of the class of two-stage estimators, in which case the latter are inefficient under model $\mathcal{M}\cap\mathcal{A}_x\cap\mathcal{A}_y$. 

Let us now consider Standard TSLS estimators assuming $\mathcal{M}^{const},\mathcal{A}^{lin}_y,\mathcal{A}^{lin}_x$
obtained by solving 
(11)-(12) with $e_x(Z,C)=(Z^T, C^T)^T$ and $e_y(Z,C)=(C^T, \alpha^{*T}_z Z+\alpha^{*T}_cC)^T$. When these models hold and moreover exposure and outcome have constant covariance matrix, conditional on $Z$ and $C$, then it is readily verified from the above expression for $d_{\rm opt}(Z,C)$
that the efficient score equations for $(\alpha^*,\beta^*,\psi^*)$ are of the form
\begin{eqnarray*}
0&=&\sum_{i=1}^n (Z_i^T, C_i^T)^T\left\{a_1\nu_i(\alpha) + a_2 \epsilon_i(\alpha,\beta,\psi)\right\}\\
0&=&\sum_{i=1}^n (C_i^T, \alpha^{*T}_z Z_i+\alpha^{*T}_cC_i)^T\left\{a_3\nu_i(\alpha) + a_4 \epsilon_i(\alpha,\beta,\psi)\right\},
\end{eqnarray*}
for constants $a_1,...,a_4$, $\nu_i(\alpha)\equiv X_i -m_x(Z_i,C_i;\alpha)$ and $\epsilon_i(\alpha,\beta,\psi)\equiv Y_i-m_y(C_i;\beta)-\psi m_x(Z_i,C_i;\alpha)$.
It can further be verified from the expression for $d_{\rm opt}(Z,C)$ above that these constants equal $a_1=\sigma_Y^2-\psi\sigma_{XY},a_2=\psi\sigma_X^2-\sigma_{XY},a_3=-\sigma_{XY}$ and $a_4=\sigma_X^2$ with $\sigma_Y^2=\mbox{Var}(Y|Z,C),\sigma_X^2=\mbox{Var}(X|Z,C)$ and $\sigma_{XY}=\mbox{Cov}(X,Y|Z,C)$. It follows that the expressions $C_i\left\{a_1\nu_i(\alpha) + a_2 \epsilon_i(\alpha,\beta,\psi)\right\}$ and $C_i\left\{a_3\nu_i(\alpha) + a_4 \epsilon_i(\alpha,\beta,\psi)\right\}$ are not collinear (unless $\mbox{Cor}(X,Y|Z,C)=1$), so that solving these equations ensures that $\sum_{i=1}^n C_i\nu_i(\alpha)=0$ and $\sum_{i=1}^n C_i\epsilon_i(\alpha,\beta,\psi)=0$. Solving the second equation in the above display is therefore equivalent to solving $0=\sum_{i=1}^n (Z_i,C_i^T)^T\left\{a_3\nu_i(\alpha) + a_4 \epsilon_i(\alpha,\beta,\psi)\right\}$. This shows that  solving the efficient score equations delivers a mathematically identical estimator as the two-stage estimator obtained by separately solving the equations 
$\sum_{i=1}^n (Z_i^T,C_i^T)^T\nu_i(\alpha)=0$ and $\sum_{i=1}^n (Z_i^T,C_i^T)^T\epsilon_i(\alpha,\beta,\psi)=0$. 
It is readily verified that this is no longer so when $m_{}(C;\psi^*)$ is a function of $C$, or when $m_x(Z,C;\alpha^*)$ is a nonlinear function of $Z$ and $C$.
$\square$
\smallskip

{\em Proof of Proposition 3.} 
Consider first the Standard TSLS estimator in model $\mathcal{M}^{const}$, which uses $e_x(Z,C)=(Z^T, C^T)^T$ and $e_y(Z,C)=(C^T,\alpha_z^TZ+\alpha^{T}_cC)^T$ (see the proof of Proposition 2).
It follows that $e_x(Z,C)$ includes a full-rank linear transformation of
$e_y(Z,C)m_{}(C;\psi)=e_y(Z,C)\psi$ for all $\psi$, hence the claimed robustness due to (14).
 The more general result for Standard TSLS estimators with $m_{}(C;\psi)$ linear in $\psi$ (i.e. $m_{}(C;\psi)=\psi^TW$ for some vector $W$, which is a function of $C$)
 is immediate upon redefining $X$ to be the multivariate exposure $XW$. $\square$

\smallskip

{\em Extension of Proposition 3.} 
Plug-In TSLS estimators of $\psi^*$ for other choices of $\mathcal{A}_x$, share the robustness of Standard TSLS under $\mathcal{M}^{const}\cap\mathcal{A}^{lin}_y$ when the population least squares residual from a regression of  $m_x(Z,C;\alpha^*)$ (with $\alpha^*$ the population limit of the OLS estimator $\hat{\alpha}$) onto $\partial m_y(C;\beta)/\partial \beta$ can be written as a linear combination of the predictors in the exposure model.

{\em Proof:} First note that for general Plug-In TSLS estimators the influence function of $\psi^*$ (in model $\mathcal{M}\cap\mathcal{A}_x\cap\mathcal{A}^{lin}_y$ with $\alpha^*$ known) is (up to a proportionality constant) equal to (12) with $e_y(Z_i,C_i)$ the population least squares residual from a regression of
 $m_x(Z_i,C_i;\alpha^*)\partial m_{}(C_i;\psi)/\partial\psi$ onto $\partial m_y(C_i;\beta)/\partial \beta$. When $m_{}(C;\psi)=\psi$, the claimed robustness is thus obtained when the population least squares residual from a regression of
 $m_x(Z_i,C_i;\alpha^*)$ onto $\partial m_y(C_i;\beta)/\partial \beta$ is a linear combination of the predictors in the exposure model. $\square$

\smallskip

{\em Illustration to extension of Proposition 3:}
For instance, suppose that $C=(1,V)^T$, $m_x(Z,C;\alpha)=\alpha_z^TZ+\alpha_c^T C$ and $m_y(C;\beta)=\beta_1^TC+\beta_2V^2$. Then the  
Plug-In TSLS estimator of $\psi^*$ in model $\mathcal{M}^{const}$ is robust to misspecification of the exposure model when $E(Z|C)$ is linear in $C$ (or in particular $Z$ is independent of $C$) for then the
population least squares residual from a regression of
 $\alpha_z^{*T}Z+\alpha_c^{*T} C$ onto $(C^T,V^2)^T$ equals $\alpha_z^{*T}E(Z|C)+\alpha_c^{*T} C$, which is a linear combination of the predictors, $Z$ and $C$, in the exposure model.

\smallskip

{\em Proof of Proposition 4:}
It follows by the same reasoning as in the Appendix of Robins and Rotnitzky (2004) and as in the first paragraph of Appendix A of these Supplemental Materials that model $\mathcal{M}\cap\mathcal{A}_y$  is the same model for the observed data as the observed data model defined by 
\begin{equation}\label{modyay}
E\left\{Y-m_{}(C;\psi^*)X|Z,C\right\}=m_y(C;\beta^*).\end{equation}
The results of Proposition 4 
now follow from general results for conditional mean models (Chamberlain, 1987). $\square$

\smallskip

{\em Remark on efficiency in the absence of an exposure model:} 
It is instructive to consider when avoiding reliance on an exposure model will or will not result in loss of efficiency compared to using a correctly specified $\mathcal{A}_x$.
The general semiparametric efficient estimator of $(\beta^*,\psi^*)$ under model $\mathcal{M}\cap\mathcal{A}_x\cap\mathcal{A}_y$ with $d_{\rm opt}(Z,C)$ the solution to equation (\ref{eelin}) of the Supplemental Materials
belongs to the class (13) 
when all working models are linear (see Proposition 3),
as well as when there happens to be no unmeasured confounding in the sense that $X\cip U \mid Z,C$ or $Y\cip U \mid X,C$. In the latter case, $\mbox{Cov}(X,Y|Z,C)=m_{}(C;\psi^*)\mbox{Var}(X|Z,C)$ and hence it follows from the expression for $d_{\rm opt}(Z,C)$ that the semiparametric efficient estimator of $(\alpha^*,\beta^*,\psi^*)$ in  model $\mathcal{M}\cap\mathcal{A}_x\cap\mathcal{A}_y$ reduces to the solution to
\begin{eqnarray*}
0&=&\sum_{i=1}^n \frac{\partial m_x(Z_i,C_i;\alpha^*)}{\partial\alpha}\frac{X_i-m_x(Z_i,C_i;\alpha)}{ \mbox{Var}(X_i|Z_i,C_i)} \\
0&=&\sum_{i=1}^n \left(\begin{array}{cc}\partial m_y(C_i;\beta^*)/\partial\beta \\
m_x(Z_i,C_i;\alpha^*)\partial m_{}(C_i;\psi^*)/\partial\psi \end{array}\right)\frac{Y_i-m_y(C_i;\beta)-m_{}(C_i;\psi)X_i}{\mbox{Var}(Y_i-m_{}(C_i;\psi^*)X_i|Z_i,C_i)}.
\end{eqnarray*}
When the degree of confounding is weak, specification of an exposure model therefore does not increase the information about $\psi^*$ in an important way. This is unsurprising for, in that case, the exposure distribution is ancillary to the effect parameter of interest. In other words, although we may gain a lot from exploiting a correctly specified $\mathcal{A}_x$ in cases where  confounding is strong and the assumptions of Proposition 3 
are not satisfied, this is not the case otherwise.

\smallskip

{\em Proof of Proposition 6: }
We show the more general result for the Plug-in TSLS estimator of $\psi^*$ and refer to Robins (2000) and Okui et al. (2012) for the result on
the Standard TSLS estimator. The result follows from the influence function (Newey, 1990) of the two-stage estimator of $\psi^*$ in $m_{}(C;\psi^*)=\psi^*$ obtained by solving (11)-(12).  It follows by Taylor series arguments (Tsiatis, 2006) that, up to a constant factor, this influence function equals
\begin{eqnarray*}
&&\left[m_x(Z,C;{\alpha}^*)-E\left\{m_x(Z,C;{\alpha}^*)C\right\}E(CC^T)^{-1}C\right]\left\{Y-m_y(C;\beta^*)-\psi^*X\right\}\\
&&=\left\{Z-E(ZC)E(CC^T)^{-1}C\right\}\left\{Y-m_y(C;\beta^*)-\psi^*X\right\};\end{eqnarray*}
where the equality is only satisfied for specific exposure models, such as $m_x(Z,C;\alpha)=\alpha_1^TZ+\alpha_2^TC$. When $E(Z|C)=\gamma^{*T} C$ for some $\gamma^*$, then $E(ZC^T)E(CC^T)^{-1}=\gamma^*$ and thus the influence function of $\psi^*$ can be seen to be proportional to
\[\left\{Z-E(Z|C)\right\}\left\{Y-m_y(C;\beta^*)-\psi^*X\right\}.\]
This has mean zero under model $\mathcal{M}$, regardless of whether $m_y(C;\beta^*)$ equals the conditional mean of $Y-m_{}(C;\psi^*)X$ given $C$. For more general $m_{}(C;\psi^*)$, one must replace $m_x(Z,C;{\alpha}^*)$ by $m_x(Z,C;{\alpha}^*)\partial m_{}(C;\psi^*)/\partial\psi$ in the above influence function. The result is then no longer true, unless $m_{}(C;\psi^*)$ is linear in $C$ and $Z$ is independent of $C$, since otherwise 
the conditional expectation of $m_x(Z,C;{\alpha}^*)\partial m_{}(C;\psi^*)/\partial\psi$, given $C$, is not linear in $C$.
$\square$




\section*{Appendix B: Empirical Efficiency Minimization}

{\em Proof of Proposition 8: }
Minimising (19) 
w.r.t.\ $\beta$ is tantamount to minimising the numerator of (19)
which amounts to solving
\begin{eqnarray*}
0&=&E\left(\left[e(Z,C;\alpha)
-E\left\{e(Z,C;\alpha)
|C\right\}\right]^2\left\{Y-m_y(C;\beta)
-m_{}(C;\psi^*)X\right\}\partial m_y(C;\beta)/\partial \beta\right).
\end{eqnarray*}
It is immediate from this expression that the solution is $\beta^*$ at model $\mathcal{A}_y$, implying that the empirical analog of this equation delivers a CAN estimator $\tilde{\beta}$ of $\beta^*$. One can likewise show that $\tilde{\alpha}$ converges in probability to some value $\alpha$. Since $\hat{\psi}(\alpha,\beta^*)$ is a CAN estimator under $\mathcal{M}^{const}\cap(\mathcal{A}_y\cup\mathcal{A}_z)$ for all $\alpha$, we conclude that $\hat{\psi}(\tilde{\alpha},\tilde{\beta})$ is CAN under $\mathcal{M}^{const}\cap(\mathcal{A}_y\cup\mathcal{A}_z)$.
 
That the proposed procedure minimises the asymptotic variance of $\hat{\psi}(\alpha,\beta)$ is immediate from the fact that (19)
equals the asymptotic variance of $\hat{\psi}(\alpha,\beta)$ when the law of $Z$ given $C$ is known and the fact that this asymptotic variance is the same when $\alpha$ and $\beta$ are substituted by estimators that converge to the values $\alpha$ and $\beta$ (at faster than $n^{1/4}$ rate). $\Box$

In the following we address some practical considerations regarding the minimization of (19)
first w.r.t.\ $\beta$  and then w.r.t.\ $\alpha$; the former leads to Corollary 7. 
Finally we address the case when the law of $Z$ given $C$ is unknown.

{\it Minimizing (19) 
w.r.t.\ $\beta$:} As we assume ${\cal M}^{const}$,  $m_{}(C_i;\psi)=\psi$; when also $e(Z,C;\alpha)=\alpha^TCZ$ and $m_y(C;\beta)=\beta^T C$, solving the above equation can be done via weighted least squares regression 
of $Y-m_{}(C;\psi^*)X$ on $C$, using weights $(\alpha^T C)^2\left\{Z-E\left(Z|C\right)\right\}^2$.
In the special case where $Y-\psi^*X$ is not just mean independent of $Z$ given $C$, but conditionally independent of $Z$ given $C$, the above expression reduces to
\[
0=E\left[(\alpha^T C)^2\mbox{Var}\left(Z|C\right)\left\{Y-\beta^T C-\psi^*X\right\}^2\partial m_y(C;\beta)/\partial \beta\right].
\]
When furthermore $\mbox{Var}\left(Z|C\right)$ and $\alpha^T C$ do not depend on $C$, then the minimiser in $\beta$ is obtained via ordinary least squares regression of $Y-m_{}(C;\psi^*)X$ on $C$. The above reasoning proves Corollary 7,
as we formalise next.

\smallskip

{\em Proof of Corollary 7:} 
This is immediate upon noting that the Standard TSLS estimator for $\psi^*$ indexing $m_{}(C;\psi^*)=\psi^*$ is the solution to an estimating equation based on estimating function
$\left\{Z-E\left(Z\right)\right\}\left(Y-\beta^T C-\psi X\right)$
when $Z\cip C$.
It now follows from the above reasoning (i.e., upon applying empirical efficiency maximisation) that when $Y-\psi^*X$ is conditionally independent of $Z$ given $C$, the optimal choice of $\beta$ is obtained via ordinary least squares regression. It follows in particular that when $Z\cip C$, the Standard TSLS estimator is always more efficient than the corresponding estimator which ignores covariates. $\square$

\smallskip

{\it Minimizing (19) 
w.r.t.\ $\alpha$:}  This  is more complicated than for $\beta$; for pedagogic purposes, we will explain it for the case where $e(Z,C)=\alpha^{*T}CZ$. Then this boils down to minimising 
\[\frac{E\left[(\alpha^{*T} C)^2\left\{Z-E\left(Z|C\right)\right\}^2\left\{Y-\beta^{*T} C-\psi^*X\right\}^2\right)}{E\left[(\alpha^{*T} C)\left\{Z-E\left(Z|C\right)\right\}X\right]^2}.\]
When $E\left\{\left(Y-\psi^*X-\beta^{*T}C\right)^2|Z,C\right\}\equiv \sigma^2$ is constant (which is for instance satisfied when the conditional variance of $Y-\psi^*X$ given $Z$ and $C$, is constant and moreover the model $\mathcal{A}_y$ is correctly specified), then upon taking derivatives w.r.t. $\alpha$, it is seen that the minimiser $\tilde{\alpha}$ is such that
\[E\left[C\left\{Z-E\left(Z|C\right)\right\}\left(X-\rho(\tilde{\alpha}^T C)\left\{Z-E\left(Z|C\right)\right\}\sigma^2\right)\right]=0\] 
and 
\[
\rho\equiv 
\frac{
E\left[(\tilde{\alpha}^T  C)\left\{Z-E\left(Z|C\right)\right\}X\right]
}{
E\left[(\tilde{\alpha}^T C)^2\left\{Z-E\left(Z|C\right)\right\}^2\sigma^2\right]
}
\]
is the population ordinary least squares coefficient of a regression of $X$ on $(\tilde{\alpha}^T C)\left\{Z-E\left(Z|C\right)\right\}\sigma^2$.
It then follows that $\alpha$ can be estimated by ordinary least squares regression of $X$ on $\left\{Z-E\left(Z|C\right)\right\}C$. 
It is indeed easily verified that the population ordinary least squares regression coefficient $\tilde{\alpha}$ of $X$ on $\left\{Z-E\left(Z|C\right)\right\}C$ solves the above identity for 
\[\rho=\frac{E\left[(\tilde{\alpha}^TC)^2\left\{Z-E\left(Z|C\right)\right\}^2\right]}{E\left[(\tilde{\alpha}^T C)^2\left\{Z-E\left(Z|C\right)\right\}^2\sigma^2\right]}=\frac{1}{\sigma^2}.\]

{\it Law of $Z|C$ unknown:}  Suppose now that the law of $Z$ given $C$ is known only up to a finite-dimensional parameter $\gamma^*$, which is estimated via maximum likelihood.
Let $S_{\gamma}$ be the score for $\gamma^*$. Then, up to a constant factor, the variance of the double-robust estimator of $\psi^*$ under model $\mathcal{M}\cap\mathcal{A}_z$
becomes
\begin{equation}\label{vars}
\frac{\mbox{Var}\left(\left[e(Z,C;\alpha)
-E\left\{e(Z,C;\alpha)
|C\right\}\right]\left\{Y-m_y(C;\beta)
-m_{}(C;\psi^*)X\right\}-\Delta S_{\gamma}\right)}{E\left(\left[e(Z,C;\alpha)
-E\left\{e(Z,C;\alpha)
|C\right\}\right]\partial m_{}(C;\psi^*)/\partial\psi X\right)^2}
\end{equation}
with 
\[
\Delta\equiv E\left[\left\{Y-m_y(C;\beta)
-m_{}(C;\psi^*)X\right\}\frac{\partial E\left\{e(Z,C;\alpha)
|C;\gamma^* \right\}}{\partial\gamma}\right]E(S_{\gamma}S_{\gamma}^T)^{-1},\] where $\Delta S_{\gamma}$ is the orthogonal projection of \[\left[e(Z,C;\alpha)
-E\left\{e(Z,C;\alpha)
|C\right\}\right]\left\{Y-m_y(C;\beta)
-m_{}(C;\psi^*)X\right\},\] 
onto $S_{\gamma}$. Minimising the variance (\ref{vars}) over $\alpha$ and $\beta$ is therefore equivalent to minimising it over $\alpha,\beta$ and $\Delta$. 
Reconsidering the working models $e(Z,C)=\alpha^T CZ$ and
$m_y(C)=\beta^T C$, and letting $E(Z|C)=\mbox{expit}(\gamma^{*T} C)$, the variance (\ref{vars}) becomes
\[
\frac{\mbox{Var}\left(\left\{Z-E\left(Z|C\right)\right\}\left\{(\alpha^T C)\left(Y-\beta^T C-\psi^*X\right)-\Delta^T C\right\}\right)}{E\left((\alpha^T C)\left\{Z-E\left(Z|C\right)\right\} X\right)^2},\]
which can be minimised along the same lines as described earlier in this section. When the outcome model $m_y(C;\beta)$ is correctly specified or procedure BR-$\beta$ is used for estimating $\beta$, then $\Delta=0$ and hence there is no need to account for the estimation of $\gamma^*$.

\section*{References}

\begin{description}
\item
 Chamberlain, G. (1987). Asymptotic efficiency in estimation with conditional moment restrictions. \textit{Journal of Econometrics}, {\bf 34}, 305-334.
\vspace*{-.25cm} \item
Newey, W.K. (1990) Semiparametric Efficiency Bounds. {\it Journal of Applied Econometrics}, {\bf 5}, 99-135. 
\vspace*{-.25cm} \item
 Okui, R., Small, D.S., Tan, Z.Q. and Robins, J.M. (2012). Doubly robust instrumental variable regression. \textit{Statistica Sinica}, {\bf 22}, 173-205.
\vspace*{-.25cm} \item
Robins JM. (2000). Robust estimation in sequentially ignorable missing data and causal inference models. \textit{Proceedings of the American Statistical Association}, Section on Bayesian Statistical Science 1999, pp. 6-10.
\vspace*{-.25cm} \item
Robins, J.M., and Rotnitzky, A.\ (2004).
Estimation of treatment effects in randomised trials with non-compliance and a dichotomous outcome using structural mean models. {\it Biometrika}, {\bf 91}, 763-783.
\end{description}